\documentclass[apj,iop]{emulateapj}
\usepackage{graphicx,xfrac,amsmath,longtable}

\tabletypesize{\scriptsize}
\newcommand{\err}[2]{\ensuremath{^{_{+#1}}_{^{-#2}}}}
\newcommand{\ee}[2]{\ensuremath{{#1}\!\times\!10^{#2}}}
\newcommand{\Msun}{\ensuremath{M_\odot}}

\newcommand{\ergcms}{\ensuremath{\mathrm{erg~cm}^{-2}~\mathrm{s}^{-1}}}
\newcommand{\chandra}{\textit{Chandra}}
\newcommand{\cxo}{\textit{Chandra X-ray Observatory}}

\newcommand{\Fx}{\mbox{$F_{\rm X}$}}
\newcommand{\Fxtot}{\mbox{$F_{\rm X,tot}$}}

\newcommand{\rein}{\ensuremath{r_\mathrm{Ein}}}

\newcommand{\hetwolens}{HE\,0230$-$2130}
\newcommand{\mglens}{MG\,J0414$+$0534}
\newcommand{\hefourlens}{HE\,0435$-$1223}
\newcommand{\rxninelens}{RX\,J0911$+$0551}
\newcommand{\sdssninelens}{SDSS\,J0924$+$0219}

\newcommand{\heelevenlens}{HE\,1113$-$0641}
\newcommand{\pglens}{PG\,1115$+$080}
\newcommand{\rxelevenlens}{RX\,J1131$-$1231}
\newcommand{\sdsselevenlens}{SDSS\,1138$+$0314}
\newcommand{\hlens}{H\,1413$+$117}
\newcommand{\blens}{B\,1422$+$231}
\newcommand{\wfitwentysixlens}{WFI\,J2026$-$4536}
\newcommand{\wfithirtythreelens}{WFI\,J2033$-$4723}
\newcommand{\qlens}{Q\,2237$+$0305}

\newcommand{\lfx}{\ensuremath{\mathcal{X}}}
\newcommand{\Xhm}{\ensuremath{\lfx_\mathrm{HM}}}
\newcommand{\Xhs}{\ensuremath{\lfx_\mathrm{HS}}}
\newcommand{\Xlm}{\ensuremath{\lfx_\mathrm{LM}}}
\newcommand{\Xls}{\ensuremath{\lfx_\mathrm{LS}}}

\begin{document}
\shorttitle{published in {\sc The Astrophysical Journal}, 744:11, 2012 January 10}
\shortauthors{Pooley et al.}
\slugcomment{published in {\sc The Astrophysical Journal}, 744:111, 2012 January 10}

\title{X-Ray And Optical Flux Ratio Anomalies In Quadruply Lensed Quasars. II. Mapping the Dark Matter Content in Elliptical Galaxies}

\author{David Pooley\altaffilmark{1,2}, Saul Rappaport\altaffilmark{3},  Jeffrey A.\ Blackburne \altaffilmark{4}, Paul  L.\ Schechter\altaffilmark{3}, and Joachim Wambsganss\altaffilmark{5}}  

\altaffiltext{1}{Eureka Scientific, Inc., 2452 Delmer Street, Suite 100, Oakland, CA 94602, USA, {\tt davepooley@me.com}}
\altaffiltext{2}{Department of Astronomy, University of Texas, Austin, TX 78712, USA}
\altaffiltext{3}{Department of Physics and Kavli Institute for Astrophysics and Space Research, Massachusetts Institute of Technology, Cambridge, MA 02139, USA}
\altaffiltext{4}{Department of Astronomy, Ohio State University, 140 West 18th Avenue, Columbus, OH 43210, USA}
\altaffiltext{5}{Astronomisches Rechen-Institut, Zentrum f\"ur Astronomie der Universit\"at Heidelberg, Moenchhofstr.12-14, 69120 Heidelberg, Germany}


\begin{abstract}
We present a microlensing analysis of 61 \chandra\ observations of 14 quadruply lensed quasars.  X-ray flux measurements of the individual quasar images give a clean determination of the microlensing effects in the lensing galaxy and thus offer a direct assessment of the local fraction of stellar matter making up the total integrated mass along the lines of sight through the lensing galaxy.  A Bayesian analysis of the ensemble of lensing galaxies gives a most likely local stellar fraction of 7\%, with the other 93\% in a smooth, dark matter component, at a mean impact parameter $R_c$ of 6.6 kpc from the center of the lensing galaxy.  We divide the systems into smaller ensembles based on $R_c$ and find that the most likely local stellar fraction varies qualitatively and quantitatively as expected, decreasing as a function of $R_c$.

\end{abstract}

\section{Introduction}
Through decades of study on quadruply gravitationally lensed quasars, it has been well established that simple mass models of lensing galaxies---a monopole plus a quadrupole---are fairly successful in describing the overall surface density of matter in a lensing galaxy.  However, these models give no indication of the type of matter present, whether it is in a smooth, dark matter component,  in a clumpy component like stars or dark matter sub-halos, or in some combination of the two \citep[e.g.,][]{2006ApJ...640...47K}.

The mass models are smooth by design, but it has become clear that some small-scale structure, i.e., some clumpiness, must be present in the lensing galaxies.  Much of the observational evidence for this comes from the ``flux ratio anomalies'' seen in several systems, in which the simple mass model correctly predicts the locations of the quasar images but fails on the relative fluxes of those images \citep[e.g.,][]{2002ApJ...567L...5M,2004ApJ...610...69K,2007ApJ...661...19P}.  The small-scale structure further lenses the background quasar with little effect on the positions of the images but large effect on their brightness. 

Arguments were put forth for two leading candidates (stars or dark matter halos) that might constitute this small-scale structure.  In the case of {\em milli}lensing, dark matter condensations of $10^4$--$10^6$~\Msun\ are responsible \citep{1992ApJ...397L...1W, 1995ApJ...443...18W, 1998MNRAS.295..587M, 2001ApJ...563....9M, 2002ApJ...572...25D, 2002ApJ...565...17C}, whereas in the case of {\em micro}lensing, stars in the lensing galaxy are responsible \citep{1995ApJ...443...18W, 2002ApJ...580..685S}.  

Our previous study of flux ratio anomalies in 10 systems \citep[Paper I;][]{2007ApJ...661...19P} as well as studies of individual lenses such as \rxelevenlens\ \citep{2006ApJ...640..569B, 2007ASPC..371...43K, 2009ApJ...693..174C} and \pglens\ \citep{2006ApJ...648...67P, 2009ApJ...697.1892P, 2008ApJ...689..755M} provide very strong evidence that microlensing is the primary cause of the flux ratio anomalies.  We have shown that the flux ratios are more anomalous in X-rays than at optical wavelengths.   This is because the optical emitting region of the quasar accretion disk is comparable in angular size to the Einstein radii of the microlensing stars while the X-ray emitting region is considerably smaller.  If millilensing were responsible for the anomalies, there should be no chromatic effect between X-rays and optical \citep[contrary to what is seen; see also][]{2011ApJ...729...34B} since both regions would be essentially point sources compared to the Einstein radius of a dark matter sub-halo. In addition, we would not expect temporal variation of the flux ratios within a human lifetime in the case of millilensing, whereas they are naturally expected to vary on timescales of months to years in the case of microlensing (as is indeed observed).

Because the X-rays come from a region much smaller than the Einstein radii of the microlensing stars, they offer a much cleaner signal of microlensing than what is available from the optical, which gives a convolution of microlensing and the finite size of the optical emitting region of the quasar.  

\citet{2002ApJ...580..685S} explored the microlensing effects of different fractional contributions of stars and dark matter to the total surface density, especially in regard to the probability of strong observable microlensing effects on saddle point images.  They found that the probability of a strong demagnification of a saddle point image, which is often seen in the observations, was relatively low for stellar fractions of 2\% and 100\% but became appreciable for stellar fractions of 5\%--25\% (see, e.g., their Figure 3).  They exploited this finding to determine the most likely stellar fraction for an ensemble of 11 lensing galaxies at the typical impact parameter of image formation \citep{2004IAUS..220..103S}.

They noted, however, that their analysis produced inconsistent results unless they assumed that the optical continuum emitting regions had an extended component. As we showed in Paper I, this is indeed the case, and it complicates the use of the optical data for dark matter determinations.  The X-rays, coming from essentially a point source region as far as the microlenses are concerned, do not suffer such complications and offer a much more promising avenue.

In \citet{2009ApJ...697.1892P}, we applied the technique of \citet{2004IAUS..220..103S} to \cxo\ observations of \pglens\ and constrained the dark matter fraction to $\sim$80\%--95\% at a characteristic distance of $\sim$6 kpc from the center of the lensing galaxy.  In this work, we extend that analysis to \chandra\ observations of 14 gravitational lenses: \hetwolens\ (1 obs.), \mglens\ (7 obs.), \hefourlens\ (1 obs.), \rxninelens\ (2 obs.), \sdssninelens\ (1 obs.), \heelevenlens\ (1 obs.), \pglens\ (6 obs.), \rxelevenlens\ (22 obs.), \sdsselevenlens\ (1 obs.), \hlens\ (2 obs.), \blens\ (3 obs.), \wfitwentysixlens\ (1 obs.), \wfithirtythreelens\ (1 obs.), and \qlens\ (12 obs.).  The observations and data reduction are described in Section \ref{sec:xray-obs}.  Our analysis of the X-ray data to obtain fluxes for each of the four images in each observation is presented in Sections \ref{sec:spectralanalysis} and \ref{sec:imageanalysis}.  The Bayesian microlensing analysis is given in Section \ref{sec:dm}, and we discuss the results in Sections \ref{sec:discuss}.  We summarize our findings in Section \ref{sec:summary}.

\section{Observations and Data Reduction}
\label{sec:xray-obs}
We utilize publicly available \chandra\ observations of 14 X-ray bright quadruply lensed quasars.  All data were downloaded from the \chandra\ archive, and reduction was performed using the \chandra\ Interactive Analysis of Observations (CIAO) software, version 4.2.   All observations were taken with the telescope aimpoint on the Advanced CCD Imaging Spectrometer (ACIS) S3 chip. The data were reprocessed using the CALDB\,4.3.0 set of calibration files (gain maps, quantum efficiency, quantum efficiency uniformity, effective area) including a new bad pixel list made with the {\tt acis\_run\_hotpix} tool.  The reprocessing was done without including the pixel randomization that is added during standard processing.  This omission slightly improves the point-spread function (PSF).  The data were filtered using the standard {\it ASCA} grades and excluding both bad pixels and software-flagged cosmic-ray events. Intervals of strong background flaring were searched for, and a few were found. In all cases, the flares were mild enough that removing the intervals would have decreased the signal to noise of the quasar images since it would have removed substantially more source flux than background flux within the small extraction regions.  Therefore, we did not remove any flaring intervals. The observation IDs, dates of observation, and exposure times are given in Table~\ref{tab:xrayfluxes}.

\section{Analysis of X-ray Spectra}
\label{sec:spectralanalysis}
For each observation of each system, we extracted events in large regions which enclosed all four images of the quasar.  We fit the spectra of these events to determine the total flux \Fxtot\ detected in each observation. Later, as we describe in Section \ref{sec:imageanalysis}, we perform two-dimensional image fitting to determine what fractions of \Fxtot\ to assign to individual images.

The source extraction regions were 4\farcs92 in radius, and we extracted background counts from an annulus around each system with an inner radius of 7\farcs38 and an outer radius of 14\farcs76.  For each observation, we simultaneously fit the source and background spectra in Sherpa 4.2 \citep{2001SPIE.4477...76F} using modified \citet{1979ApJ...228..939C} statistics (``cstat'' in Sherpa) and the \citet{nm65} optimization method (``simplex'' in Sherpa).  Both source and background were modeled as absorbed, independent power laws.  The absorption column density was fixed at the Galactic value in the direction of the lens based on the maps of \citet{1990ARA&A..28..215D}.  These simple spectral models are meant only to reproduce the gross X-ray spectral shape for flux estimation and not to test for the presence of additional features such as extragalactic absorption or spectral emission lines; nonetheless, the reduced fit statistics indicate more than adequate agreement between the simple models and the X-ray data. Fits were performed over the 0.5--8 keV energy range.  The best fit power-law index, reduced cstat statistic, and unabsorbed X-ray flux are reported in Table~\ref{tab:xrayfluxes} for each observation.

To calculate the uncertainties in the X-ray fluxes, we used the Sherpa tool ``sample\_energy\_flux.''  This tool used 1000 samples of the power-law index and amplitude from their normal distributions to calculate 1000 values of the 0.5--8 keV flux.  The standard deviation of that flux sample is reported as the uncertainty on the X-ray flux in Table~\ref{tab:xrayfluxes}.

\section{Analysis of X-ray Images}
\label{sec:imageanalysis}
As in our previous work \citep{2006ApJ...648...67P,2007ApJ...661...19P,2009ApJ...697.1892P}, we rely on two-dimensional fitting to determine the relative X-ray intensities of the four quasar images in each system.  In this work, we have explored several strategies to achieve the best determinations of these relative intensities.

First, we make three sky images of each observation at a resolution of  0\farcs0492 per pixel using events in the 0.3--8.0 keV energy range.  The first image comprises all events in the reprocessed Level 2 event list, and we refer to this as a ``standard'' image.  The second and third images are made after applying the subpixel event repositioning (SER) algorithm of \citet{2004ApJ...610.1204L} and either including all split-pixel events or including only the corner-split events.  We refer to these images as the ``SER'' image and the ``SER-CO'' image, respectively.  These SER images were made with the aim to improve the determination of the X-ray intensities by improving the effective \chandra\ spatial resolution.  Their efficacy is discussed below.  The main idea behind the SER method is that a charge cloud split between two or more CCD pixels can be better positioned based on the distribution of charge among the pixels.  The corner-split events provide a more precise repositioning of the event but are much fewer in number (see Section \ref{sec:sercomp}).

The models we use to fit the images consist of a fixed background level determined from a large, source-free area near the system plus four other components to represent the four quasar images. In all cases, the four components are fixed in their relative positions to each other, with the absolute position allowed to vary to the best fit location.   Previously, we used two-dimensional Gaussians for these components, with each of the four Gaussians in a fit constrained to have the same full width at half-maximum (FWHM) but allowing the specific value of the FWHM to vary to the best-fit value.   We repeat that analysis using Gaussians, and we now also explore other choices to model the quasar images.    Our aim is to find the best representation of the X-ray data that yields the smallest uncertainties in the model amplitudes, i.e., the individual fluxes of the four quasar images.

\LongTables
\setlength{\tabcolsep}{2pt}
\begin{deluxetable*}{lllrrrrrrrcc}[t!]
\tablewidth{0pt}
\tablecaption{\chandra\ Observations of 14 Quadruply Lensed Quasars}
\tablehead{\colhead{System Number}& \colhead{System Name} & \colhead{Date} & \colhead{ObsId} & \colhead{Exp.} & \colhead{HM Fraction} & \colhead{HS Fraction} &\colhead{LM Fraction} & \colhead{LS Fraction} & \colhead{Total \Fx/$10^{-13}$} & \colhead{PL} & \colhead{Reduc.} \\ & & & & (ks) & & & & & \colhead{(\ergcms)}  & \colhead{Ind.} & \colhead{C-stat} }
\startdata
1&          \hetwolens & 2000 Oct 14 & \dataset[ADS/Sa.CXO\#obs/01642]{ 1642} & 14.8 & 0.494\err{0.074}{0.064} & 0.168\err{0.034}{0.029} & 0.245\err{0.041}{0.035} & 0.093\err{0.021}{0.018} &  2.59 $\pm$  0.55 & 1.8 & 0.82\\[1.5ex]
2&             \mglens & 2000 Jan 13 & \dataset[ADS/Sa.CXO\#obs/00417]{  417} &  6.6 & 0.428\err{0.077}{0.070} & 0.337\err{0.079}{0.071} & 0.190\err{0.034}{0.029} & 0.045\err{0.014}{0.012} &  7.04 $\pm$  1.22 & 1.0 & 0.56\\
&             \mglens & 2000 Apr 02 & \dataset[ADS/Sa.CXO\#obs/00418]{  418} &  7.4 & 0.421\err{0.072}{0.063} & 0.282\err{0.063}{0.056} & 0.185\err{0.034}{0.029} & 0.111\err{0.024}{0.020} &  7.38 $\pm$  1.28 & 1.0 & 0.78\\
&             \mglens & 2000 Aug 16 & \dataset[ADS/Sa.CXO\#obs/00421]{  421} &  7.3 & 0.455\err{0.088}{0.074} & 0.271\err{0.067}{0.058} & 0.196\err{0.040}{0.033} & 0.077\err{0.020}{0.016} &  6.98 $\pm$  1.19 & 1.1 & 0.55\\
&             \mglens & 2000 Nov 16 & \dataset[ADS/Sa.CXO\#obs/00422]{  422} &  7.5 & 0.460\err{0.075}{0.068} & 0.284\err{0.062}{0.057} & 0.140\err{0.025}{0.022} & 0.116\err{0.022}{0.019} &  8.39 $\pm$  1.38 & 1.0 & 0.56\\
&             \mglens & 2001 Feb 05 & \dataset[ADS/Sa.CXO\#obs/01628]{ 1628} &  9.0 & 0.389\err{0.061}{0.056} & 0.340\err{0.062}{0.057} & 0.206\err{0.030}{0.026} & 0.065\err{0.014}{0.012} &  8.08 $\pm$  1.21 & 1.1 & 0.57\\
&             \mglens & 2001 Nov 09 & \dataset[ADS/Sa.CXO\#obs/03395]{ 3395} & 28.4 & 0.414\err{0.041}{0.038} & 0.305\err{0.035}{0.033} & 0.182\err{0.018}{0.016} & 0.099\err{0.012}{0.011} &  6.74 $\pm$  0.63 & 1.1 & 0.91\\
&             \mglens & 2002 Jan 08 & \dataset[ADS/Sa.CXO\#obs/03419]{ 3419} & 96.7 & 0.411\err{0.020}{0.019} & 0.312\err{0.019}{0.018} & 0.197\err{0.009}{0.009} & 0.079\err{0.005}{0.005} &  6.79 $\pm$  0.33 & 1.1 & 1.42\\[1.5ex]
3&         \hefourlens & 2006 Dec 17 & \dataset[ADS/Sa.CXO\#obs/07761]{ 7761} & 10.0 & 0.167\err{0.029}{0.025} & 0.183\err{0.031}{0.027} & 0.480\err{0.064}{0.056} & 0.170\err{0.029}{0.025} &  3.76 $\pm$  0.89 & 1.9 & 0.44\\[1.5ex]
4&         \rxninelens & 1999 Nov 02 & \dataset[ADS/Sa.CXO\#obs/00419]{  419} & 28.8 & 0.216\err{0.062}{0.053} & 0.563\err{0.101}{0.086} & 0.171\err{0.037}{0.031} & 0.050\err{0.023}{0.018} &  1.34 $\pm$  0.46 & 1.1 & 0.74\\
&         \rxninelens & 2000 Oct 29 & \dataset[ADS/Sa.CXO\#obs/01629]{ 1629} &  9.8 & 0.086\err{0.122}{0.082} & 0.583\err{0.207}{0.148} & 0.200\err{0.093}{0.063} & 0.132\err{0.080}{0.054} &  1.24 $\pm$  0.99 & 1.0 & 0.36\\[1.5ex]
5&       \sdssninelens & 2005 Feb 24 & \dataset[ADS/Sa.CXO\#obs/05604]{ 5604} & 17.9 & 0.727\err{0.186}{0.145} & 0.021\err{0.044}{0.032} & 0.168\err{0.060}{0.046} & 0.084\err{0.043}{0.031} &  0.47 $\pm$  0.21 & 2.2 & 0.33\\[1.5ex]
6&       \heelevenlens & 2007 Jan 28 & \dataset[ADS/Sa.CXO\#obs/07760]{ 7760} & 15.0 & 0.168\err{0.130}{0.151} & 0.316\err{0.104}{0.098} & 0.447\err{0.111}{0.122} & 0.068\err{0.101}{0.100} &  2.04 $\pm$  0.36 & 2.2 & 0.47\\[1.5ex]
7&             \pglens & 2000 Jun 02 & \dataset[ADS/Sa.CXO\#obs/00363]{  363} & 26.5 & 0.585\err{0.043}{0.040} & 0.112\err{0.023}{0.021} & 0.150\err{0.015}{0.014} & 0.154\err{0.015}{0.014} &  4.80 $\pm$  0.56 & 1.6 & 0.72\\
&             \pglens & 2000 Nov 03 & \dataset[ADS/Sa.CXO\#obs/01630]{ 1630} &  9.8 & 0.659\err{0.081}{0.072} & 0.107\err{0.032}{0.028} & 0.126\err{0.022}{0.019} & 0.107\err{0.020}{0.017} &  5.16 $\pm$  0.94 & 1.7 & 0.49\\
&             \pglens & 2008 Jan 31 & \dataset[ADS/Sa.CXO\#obs/07757]{ 7757} & 28.8 & 0.413\err{0.030}{0.028} & 0.342\err{0.028}{0.027} & 0.120\err{0.010}{0.009} & 0.125\err{0.010}{0.010} &  7.11 $\pm$  0.67 & 1.7 & 0.71\\
&             \pglens & 2008 Nov 02 & \dataset[ADS/Sa.CXO\#obs/10730]{10730} & 14.6 & 0.348\err{0.076}{0.069} & 0.401\err{0.078}{0.070} & 0.145\err{0.029}{0.024} & 0.106\err{0.023}{0.020} &  3.63 $\pm$  0.86 & 1.3 & 0.55\\
&             \pglens & 2009 Feb 09 & \dataset[ADS/Sa.CXO\#obs/10795]{10795} & 14.5 & 0.394\err{0.071}{0.064} & 0.396\err{0.071}{0.064} & 0.099\err{0.020}{0.017} & 0.110\err{0.022}{0.019} &  4.05 $\pm$  0.84 & 1.3 & 0.59\\
&             \pglens & 2009 Mar 27 & \dataset[ADS/Sa.CXO\#obs/10796]{10796} & 14.6 & 0.335\err{0.053}{0.048} & 0.475\err{0.066}{0.059} & 0.098\err{0.018}{0.015} & 0.092\err{0.017}{0.015} &  5.54 $\pm$  1.00 & 1.3 & 0.67\\[1.5ex]
8&       \rxelevenlens & 2004 Apr 12 & \dataset[ADS/Sa.CXO\#obs/04814]{ 4814} & 10.0 & 0.634\err{0.035}{0.033} & 0.087\err{0.009}{0.008} & 0.220\err{0.015}{0.014} & 0.059\err{0.006}{0.006} & 20.20 $\pm$  1.36 & 1.5 & 0.71\\
&       \rxelevenlens & 2006 Mar 10 & \dataset[ADS/Sa.CXO\#obs/06913]{ 6913} &  4.9 & 0.501\err{0.043}{0.039} & 0.264\err{0.027}{0.025} & 0.163\err{0.018}{0.017} & 0.071\err{0.011}{0.009} & 19.56 $\pm$  1.68 & 1.7 & 0.55\\
&       \rxelevenlens & 2006 Mar 15 & \dataset[ADS/Sa.CXO\#obs/06912]{ 6912} &  4.4 & 0.476\err{0.044}{0.040} & 0.279\err{0.030}{0.027} & 0.169\err{0.020}{0.018} & 0.076\err{0.011}{0.010} & 20.45 $\pm$  1.86 & 1.7 & 0.56\\
&       \rxelevenlens & 2006 Apr 15 & \dataset[ADS/Sa.CXO\#obs/06914]{ 6914} &  4.9 & 0.418\err{0.041}{0.037} & 0.350\err{0.037}{0.033} & 0.118\err{0.017}{0.015} & 0.113\err{0.015}{0.014} & 17.73 $\pm$  1.77 & 1.6 & 0.49\\
&       \rxelevenlens & 2006 Nov 10 & \dataset[ADS/Sa.CXO\#obs/06915]{ 6915} &  4.8 & 0.262\err{0.014}{0.013} & 0.624\err{0.029}{0.027} & 0.083\err{0.007}{0.006} & 0.030\err{0.003}{0.003} & 51.37 $\pm$  2.74 & 1.8 & 0.74\\
&       \rxelevenlens & 2006 Nov 13 & \dataset[ADS/Sa.CXO\#obs/06916]{ 6916} &  4.8 & 0.272\err{0.014}{0.014} & 0.627\err{0.028}{0.027} & 0.083\err{0.007}{0.006} & 0.018\err{0.003}{0.002} & 53.06 $\pm$  2.72 & 1.8 & 0.78\\
&       \rxelevenlens & 2006 Dec 17 & \dataset[ADS/Sa.CXO\#obs/07786]{ 7786} &  4.9 & 0.279\err{0.015}{0.014} & 0.605\err{0.029}{0.028} & 0.088\err{0.007}{0.007} & 0.028\err{0.003}{0.003} & 49.08 $\pm$  2.69 & 1.7 & 0.72\\
&       \rxelevenlens & 2007 Jan 01 & \dataset[ADS/Sa.CXO\#obs/07785]{ 7785} &  4.7 & 0.278\err{0.017}{0.016} & 0.607\err{0.033}{0.031} & 0.080\err{0.008}{0.007} & 0.035\err{0.004}{0.004} & 45.30 $\pm$  2.60 & 1.8 & 0.70\\
&       \rxelevenlens & 2007 Feb 13 & \dataset[ADS/Sa.CXO\#obs/07787]{ 7787} &  4.7 & 0.300\err{0.018}{0.017} & 0.569\err{0.030}{0.029} & 0.085\err{0.008}{0.007} & 0.046\err{0.005}{0.005} & 47.69 $\pm$  2.55 & 1.8 & 0.65\\
&       \rxelevenlens & 2007 Feb 18 & \dataset[ADS/Sa.CXO\#obs/07788]{ 7788} &  4.4 & 0.272\err{0.018}{0.017} & 0.607\err{0.034}{0.032} & 0.080\err{0.008}{0.008} & 0.041\err{0.005}{0.005} & 43.82 $\pm$  2.52 & 1.8 & 0.59\\
&       \rxelevenlens & 2007 Apr 16 & \dataset[ADS/Sa.CXO\#obs/07789]{ 7789} &  4.7 & 0.295\err{0.018}{0.017} & 0.587\err{0.031}{0.030} & 0.087\err{0.008}{0.007} & 0.031\err{0.004}{0.004} & 46.85 $\pm$  2.60 & 1.8 & 0.64\\
&       \rxelevenlens & 2007 Apr 25 & \dataset[ADS/Sa.CXO\#obs/07790]{ 7790} &  4.7 & 0.309\err{0.019}{0.018} & 0.537\err{0.031}{0.029} & 0.106\err{0.009}{0.009} & 0.048\err{0.005}{0.005} & 42.94 $\pm$  2.56 & 1.8 & 0.68\\
&       \rxelevenlens & 2007 Jun 04 & \dataset[ADS/Sa.CXO\#obs/07791]{ 7791} &  4.7 & 0.347\err{0.020}{0.019} & 0.525\err{0.028}{0.026} & 0.090\err{0.008}{0.007} & 0.039\err{0.005}{0.004} & 44.71 $\pm$  2.45 & 1.9 & 0.74\\
&       \rxelevenlens & 2007 Jun 11 & \dataset[ADS/Sa.CXO\#obs/07792]{ 7792} &  4.7 & 0.356\err{0.020}{0.019} & 0.527\err{0.028}{0.026} & 0.083\err{0.008}{0.007} & 0.035\err{0.004}{0.004} & 45.17 $\pm$  2.39 & 1.9 & 0.69\\
&       \rxelevenlens & 2007 Jul 24 & \dataset[ADS/Sa.CXO\#obs/07793]{ 7793} &  4.7 & 0.346\err{0.021}{0.019} & 0.524\err{0.029}{0.027} & 0.103\err{0.009}{0.008} & 0.027\err{0.004}{0.004} & 43.93 $\pm$  2.53 & 1.8 & 0.63\\
&       \rxelevenlens & 2007 Jul 30 & \dataset[ADS/Sa.CXO\#obs/07794]{ 7794} &  4.7 & 0.339\err{0.017}{0.017} & 0.529\err{0.025}{0.024} & 0.106\err{0.008}{0.007} & 0.026\err{0.003}{0.003} & 57.93 $\pm$  2.92 & 1.8 & 0.74\\
&       \rxelevenlens & 2008 Mar 16 & \dataset[ADS/Sa.CXO\#obs/09180]{ 9180} & 14.3 & 0.355\err{0.012}{0.012} & 0.476\err{0.015}{0.015} & 0.127\err{0.006}{0.006} & 0.042\err{0.003}{0.003} & 43.47 $\pm$  1.48 & 1.7 & 0.84\\
&       \rxelevenlens & 2008 Apr 13 & \dataset[ADS/Sa.CXO\#obs/09181]{ 9181} & 14.3 & 0.354\err{0.010}{0.010} & 0.509\err{0.014}{0.014} & 0.105\err{0.004}{0.004} & 0.033\err{0.002}{0.002} & 52.51 $\pm$  1.66 & 1.7 & 0.92\\
&       \rxelevenlens & 2008 Apr 23 & \dataset[ADS/Sa.CXO\#obs/09237]{ 9237} & 14.3 & 0.353\err{0.011}{0.011} & 0.517\err{0.015}{0.015} & 0.106\err{0.005}{0.005} & 0.024\err{0.002}{0.002} & 48.11 $\pm$  1.60 & 1.7 & 0.84\\
&       \rxelevenlens & 2008 Jun 01 & \dataset[ADS/Sa.CXO\#obs/09238]{ 9238} & 14.2 & 0.337\err{0.013}{0.012} & 0.496\err{0.018}{0.017} & 0.100\err{0.006}{0.005} & 0.068\err{0.004}{0.004} & 36.25 $\pm$  1.38 & 1.7 & 0.78\\
&       \rxelevenlens & 2008 Jul 05 & \dataset[ADS/Sa.CXO\#obs/09239]{ 9239} & 14.3 & 0.318\err{0.011}{0.011} & 0.502\err{0.016}{0.016} & 0.101\err{0.005}{0.005} & 0.079\err{0.004}{0.004} & 41.61 $\pm$  1.48 & 1.7 & 0.83\\
&       \rxelevenlens & 2008 Nov 11 & \dataset[ADS/Sa.CXO\#obs/09240]{ 9240} & 14.3 & 0.327\err{0.012}{0.012} & 0.509\err{0.018}{0.017} & 0.101\err{0.005}{0.005} & 0.063\err{0.004}{0.004} & 38.88 $\pm$  1.50 & 1.7 & 0.78\\[1.5ex]
9&     \sdsselevenlens & 2007 Feb 13 & \dataset[ADS/Sa.CXO\#obs/07759]{ 7759} & 18.8 & 0.000\err{0.048}{0} & 0.767\err{0.173}{0.163} & 0.206\err{0.079}{0.110} & 0.027\err{0.043}{0.031} &  1.04 $\pm$  0.50 & 0.9 & 0.47\\[1.5ex]
10&              \hlens & 2000 Apr 19 & \dataset[ADS/Sa.CXO\#obs/00930]{  930} & 38.2 & 0.256\err{0.067}{0.056} & 0.477\err{0.098}{0.082} & 0.114\err{0.042}{0.034} & 0.152\err{0.046}{0.037} &  1.22 $\pm$  0.47 & 0.4 & 0.84\\
&              \hlens & 2005 Mar 30 & \dataset[ADS/Sa.CXO\#obs/05645]{ 5645} & 88.9 & 0.300\err{0.056}{0.048} & 0.271\err{0.052}{0.045} & 0.231\err{0.045}{0.039} & 0.197\err{0.040}{0.034} &  0.90 $\pm$  0.24 & 0.5 & 0.84\\[1.5ex]
11&              \blens & 2000 Jun 01 & \dataset[ADS/Sa.CXO\#obs/00367]{  367} & 28.4 & 0.425\err{0.026}{0.024} & 0.313\err{0.022}{0.021} & 0.247\err{0.016}{0.015} & 0.015\err{0.003}{0.003} & 10.02 $\pm$  1.42 & 1.4 & 0.89\\
&              \blens & 2001 May 21 & \dataset[ADS/Sa.CXO\#obs/01631]{ 1631} & 10.7 & 0.414\err{0.043}{0.040} & 0.307\err{0.037}{0.034} & 0.267\err{0.028}{0.026} & 0.012\err{0.006}{0.005} & 10.56 $\pm$  2.43 & 1.5 & 0.63\\
&              \blens & 2004 Dec 01 & \dataset[ADS/Sa.CXO\#obs/04939]{ 4939} & 47.7 & 0.401\err{0.021}{0.021} & 0.295\err{0.020}{0.019} & 0.287\err{0.015}{0.014} & 0.017\err{0.003}{0.003} &  9.45 $\pm$  1.11 & 1.6 & 0.91\\[1.5ex]
12&   \wfitwentysixlens & 2007 Jun 28 & \dataset[ADS/Sa.CXO\#obs/07758]{ 7758} & 10.0 & 0.461\err{0.122}{0.121} & 0.405\err{0.121}{0.135} & 0.104\err{0.019}{0.017} & 0.030\err{0.014}{0.012} &  5.00 $\pm$  1.05 & 2.0 & 0.49\\[1.5ex]
13& \wfithirtythreelens & 2005 Mar 10 & \dataset[ADS/Sa.CXO\#obs/05603]{ 5603} & 15.4 & 0.192\err{0.050}{0.042} & 0.322\err{0.069}{0.057} & 0.276\err{0.057}{0.047} & 0.209\err{0.047}{0.039} &  1.33 $\pm$  0.35 & 2.1 & 0.44\\[1.5ex]
14&              \qlens & 2000 Sep 06 & \dataset[ADS/Sa.CXO\#obs/00431]{  431} & 30.3 & 0.580\err{0.039}{0.036} & 0.089\err{0.011}{0.010} & 0.106\err{0.011}{0.010} & 0.225\err{0.018}{0.017} &  5.32 $\pm$  0.59 & 1.8 & 0.72\\
&              \qlens & 2001 Dec 08 & \dataset[ADS/Sa.CXO\#obs/01632]{ 1632} &  9.5 & 0.612\err{0.083}{0.072} & 0.106\err{0.025}{0.021} & 0.100\err{0.022}{0.019} & 0.182\err{0.032}{0.028} &  4.45 $\pm$  1.24 & 1.8 & 0.47\\
&              \qlens & 2006 Jan 09 & \dataset[ADS/Sa.CXO\#obs/06831]{ 6831} &  7.3 & 0.413\err{0.083}{0.069} & 0.143\err{0.042}{0.034} & 0.331\err{0.070}{0.058} & 0.113\err{0.034}{0.027} &  3.31 $\pm$  1.15 & 1.6 & 0.43\\
&              \qlens & 2006 May 01 & \dataset[ADS/Sa.CXO\#obs/06832]{ 6832} &  7.9 & 0.447\err{0.059}{0.052} & 0.136\err{0.027}{0.024} & 0.231\err{0.036}{0.031} & 0.186\err{0.031}{0.027} &  5.51 $\pm$  1.36 & 1.7 & 0.49\\
&              \qlens & 2006 May 27 & \dataset[ADS/Sa.CXO\#obs/06833]{ 6833} &  8.0 & 0.495\err{0.095}{0.079} & 0.090\err{0.032}{0.026} & 0.208\err{0.049}{0.040} & 0.206\err{0.050}{0.041} &  2.96 $\pm$  1.14 & 1.6 & 0.37\\
&              \qlens & 2006 Jun 25 & \dataset[ADS/Sa.CXO\#obs/06834]{ 6834} &  7.9 & 0.511\err{0.067}{0.059} & 0.117\err{0.026}{0.022} & 0.216\err{0.034}{0.030} & 0.156\err{0.028}{0.024} &  6.01 $\pm$  1.54 & 1.7 & 0.53\\
&              \qlens & 2006 Jul 21 & \dataset[ADS/Sa.CXO\#obs/06835]{ 6835} &  7.9 & 0.632\err{0.082}{0.072} & 0.113\err{0.025}{0.021} & 0.144\err{0.027}{0.023} & 0.111\err{0.023}{0.020} &  6.62 $\pm$  1.66 & 1.5 & 0.55\\
&              \qlens & 2006 Aug 17 & \dataset[ADS/Sa.CXO\#obs/06836]{ 6836} &  7.9 & 0.500\err{0.089}{0.074} & 0.137\err{0.037}{0.031} & 0.193\err{0.043}{0.035} & 0.171\err{0.039}{0.033} &  4.50 $\pm$  1.38 & 1.4 & 0.47\\
&              \qlens & 2006 Sep 16 & \dataset[ADS/Sa.CXO\#obs/06837]{ 6837} &  7.9 & 0.505\err{0.086}{0.073} & 0.144\err{0.036}{0.030} & 0.230\err{0.046}{0.039} & 0.121\err{0.030}{0.025} &  3.71 $\pm$  1.55 & 1.6 & 0.47\\
&              \qlens & 2006 Oct 09 & \dataset[ADS/Sa.CXO\#obs/06838]{ 6838} &  8.0 & 0.493\err{0.094}{0.078} & 0.138\err{0.038}{0.031} & 0.204\err{0.048}{0.039} & 0.165\err{0.041}{0.033} &  3.37 $\pm$  1.05 & 1.7 & 0.43\\
&              \qlens & 2006 Nov 29 & \dataset[ADS/Sa.CXO\#obs/06839]{ 6839} &  7.9 & 0.568\err{0.049}{0.045} & 0.116\err{0.017}{0.015} & 0.198\err{0.022}{0.020} & 0.118\err{0.016}{0.014} & 11.01 $\pm$  1.92 & 1.7 & 0.54\\
&              \qlens & 2007 Jan 14 & \dataset[ADS/Sa.CXO\#obs/06840]{ 6840} &  8.0 & 0.547\err{0.055}{0.049} & 0.129\err{0.020}{0.018} & 0.171\err{0.022}{0.020} & 0.154\err{0.021}{0.019} &  7.94 $\pm$  1.65 & 1.9 & 0.54
\enddata
\tablecomments{Columns 5--8 give the fractional contribution of each of the HM, HS, LM, and LS images (see Section \ref{sec:magmaps} for definitions) to the total measured X-ray flux (given in Column 9).  See text for details of the spectral and image fitting.}
\label{tab:xrayfluxes}
\end{deluxetable*}

Our first alternative is the $\beta$ profile, a two-dimensional Lorentzian with a varying power law of the form $I(r) = A(1+(r/r_0)^2)^{-\alpha}$, in which each of the four components is constrained to have the same $r_0$ and $\alpha$.  Our second alternative is a $\delta$ function convolved with an observation-specific, ray-traced PSF.  Because the \chandra\ PSF is both energy- and position-dependent, a separate PSF is constructed for each observation based on the exact off-axis location of the four quasar images and the measured spectrum of the system in that observation.  This information was input to the online \chandra\ Ray Tracer\footnote{\url{http://cxc.harvard.edu/soft/ChaRT/cgi-bin/www-saosac.cgi}} to produce ray traces of the telescope PSF for each observation.  These ray traces were projected onto the ACIS detector using Marx 4.5\footnote{\url{http://space.mit.edu/CXC/MARX/}} to produce images which were then used as the PSF convolution kernel for the $\delta$ function fits.  

\begin{figure}[b]
\centering
\includegraphics[width=0.47\textwidth]{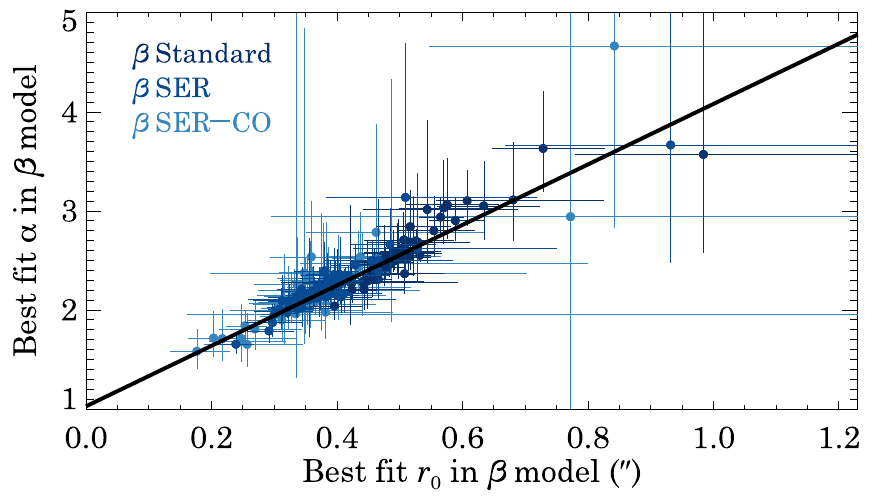}
\caption{Best-fit values of $r_0$ and $\alpha$ and 1$\sigma$ uncertainties.  Dark blue points are for the $\beta$-model fits to the standard images; medium blue is for the fits to the SER image; light blue is for the SER-CO image. The black line is a fit to the data taking into account uncertainties in both parameters.}
\label{fig:r0alpha}
\end{figure}

We therefore have a progression of models in terms of shape parameters: a zero-parameter PSF model, a one-parameter (FWHM) Gaussian model, and a two-parameter ($r_0$ and $\alpha$) $\beta$ model.  After the analysis of the $\beta$-model fits, we noticed a degeneracy between $r_0$ and $\alpha$ in all three categories of images (standard, SER, and SER-CO), shown in Figure~\ref{fig:r0alpha}.  We fit a straight line to the points, taking into account errors in both coordinates and found
\begin{equation}
\alpha = 3.05(r_0/\mathrm{arcsec}) + 1.03~~~~~.
\end{equation}
We then refit all of the images with a $\beta$ model where $\alpha$ was constrained to follow this relation (hereafter $\beta$c because it is constrained), making it essentially a one-parameter model.

\begin{figure}[b!]
\centering
\includegraphics[width=0.47\textwidth]{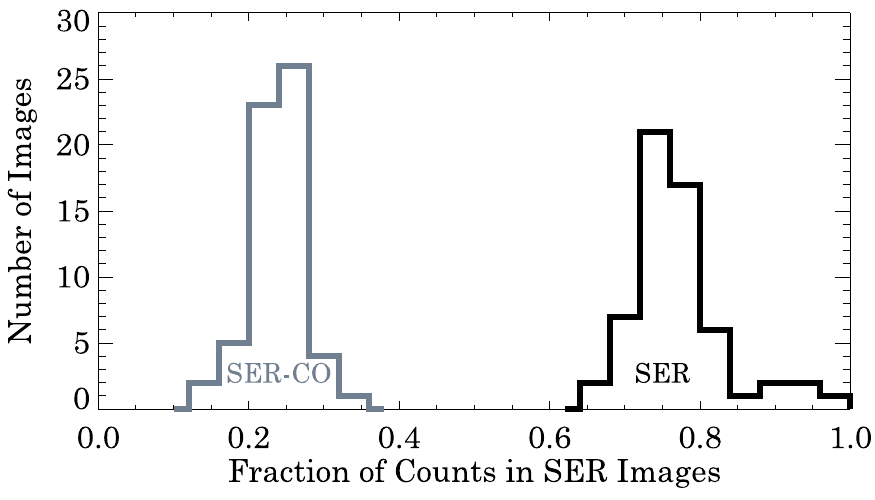}
\caption{Histograms of the fraction of counts from the standard image used in SER and SER-CO images.  The SER algorithm of \citet{2004ApJ...610.1204L} is able to use $\sim$75\% of the events on average, while the corner-only algorithm uses only $\sim$25\% on average. }
\label{fig:sercounts}
\end{figure}

We fit each of the four models to each of the three classes of images in all 61 observations for a total of 732 fits.  All fits were again performed in Sherpa 4.2 using the cstat statistic.  We employed a Monte-Carlo based optimization method (``moncar'' in Sherpa) followed up by the simplex method.

We compare the best-fit position of each of the 12 fits for an observation to test for fit fidelity.  In general, most fits agree  in  position to better than 0\farcs1, but there are some significant outliers, most common in the PSF fits, indicating a problem with those fits.  Visual inspection reveals that the fits with large position discrepancies found minima in the fit space by zeroing out one or more of the four components and shifting the other components to match up with quasar images they are not meant to represent. While these were slightly statistically better fits, they were not useful representations of the data for our purposes.  We discard all fits which were outliers of 0\farcs25 or more from the rest of the analysis, which was 14 out of the 732 fits.

\begin{figure*}[t]
\includegraphics[width=0.25\textwidth]{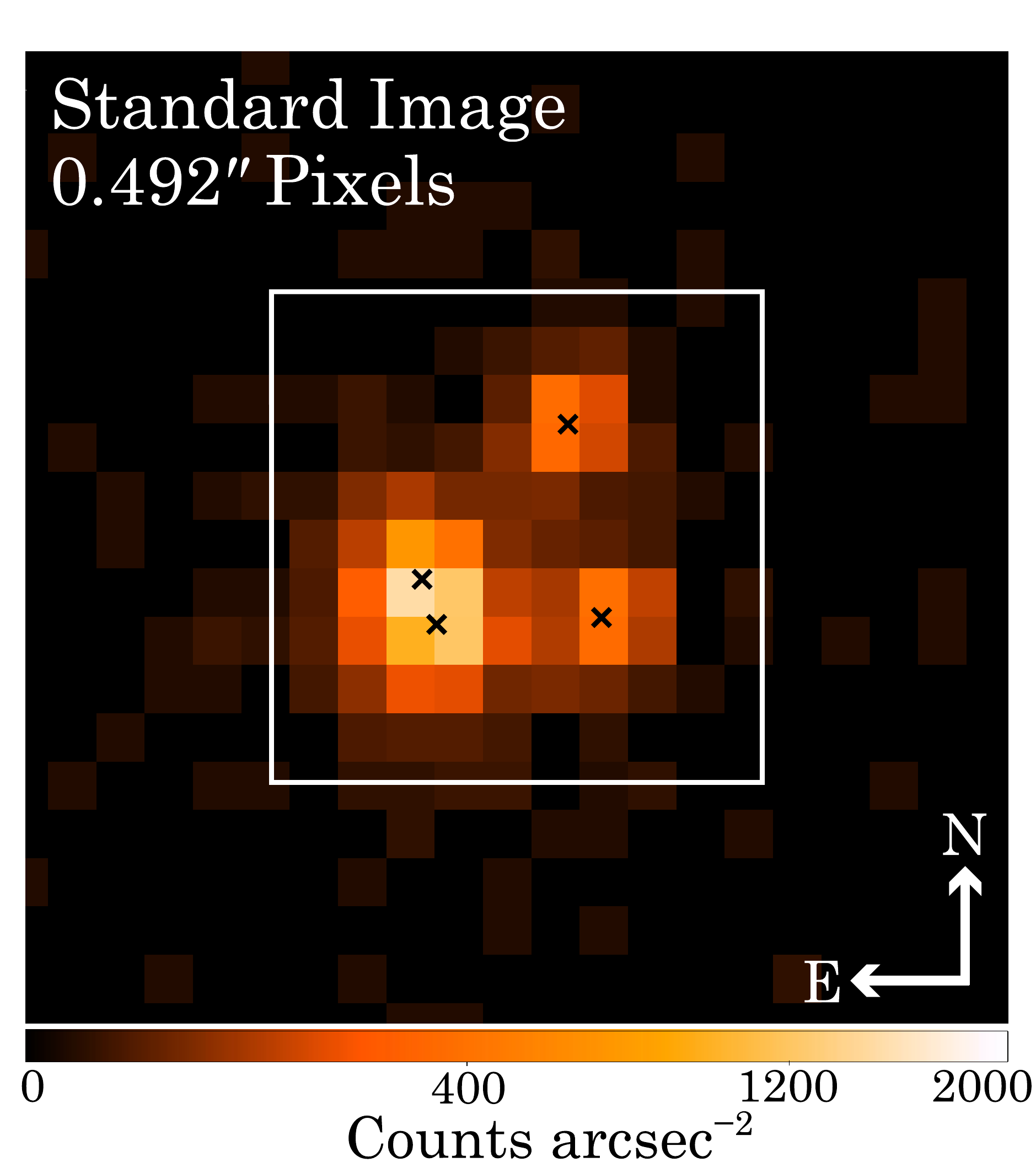}\hglue 0.0in\includegraphics[width=0.25\textwidth]{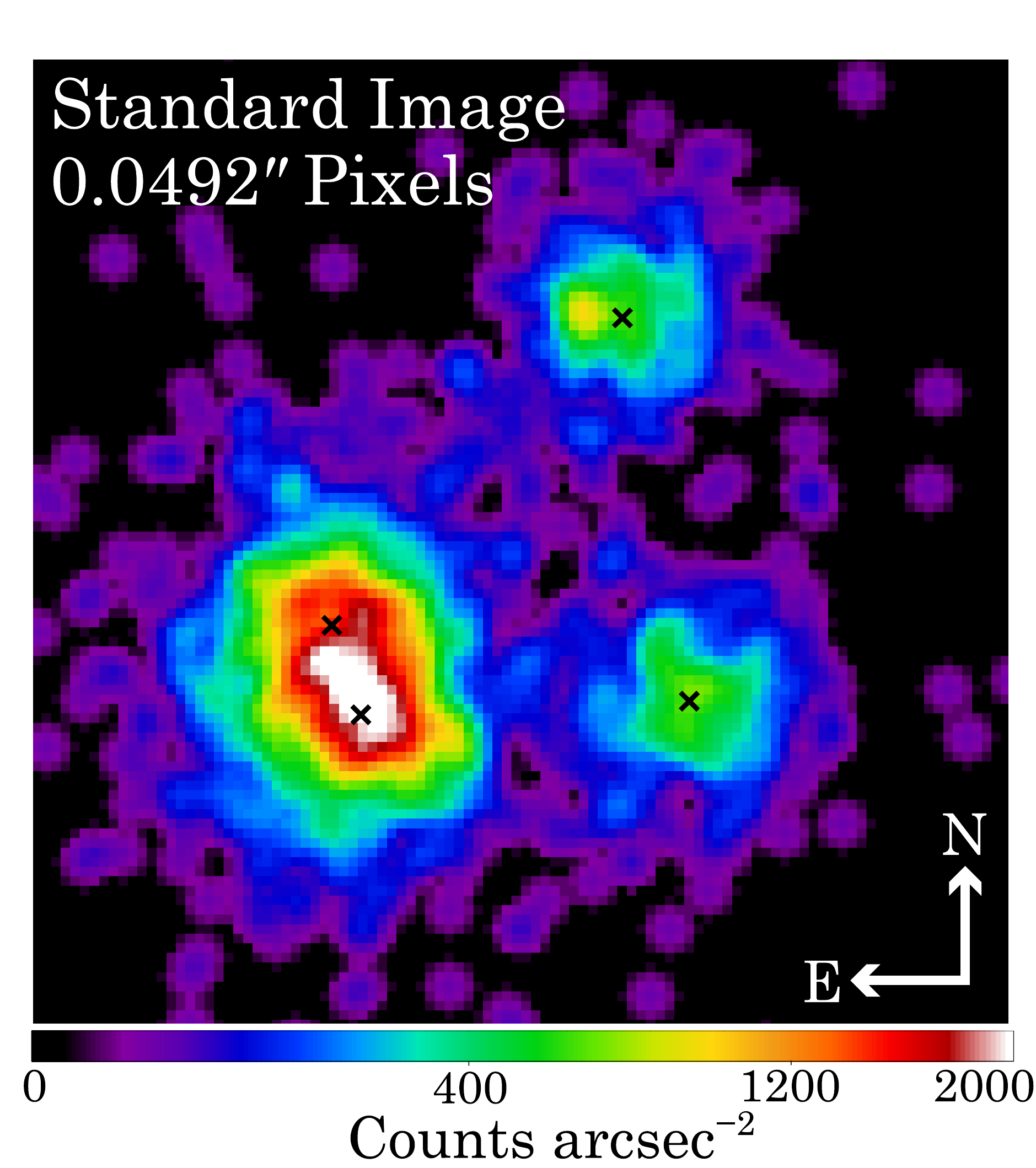}\hglue 0.0in\includegraphics[width=0.25\textwidth]{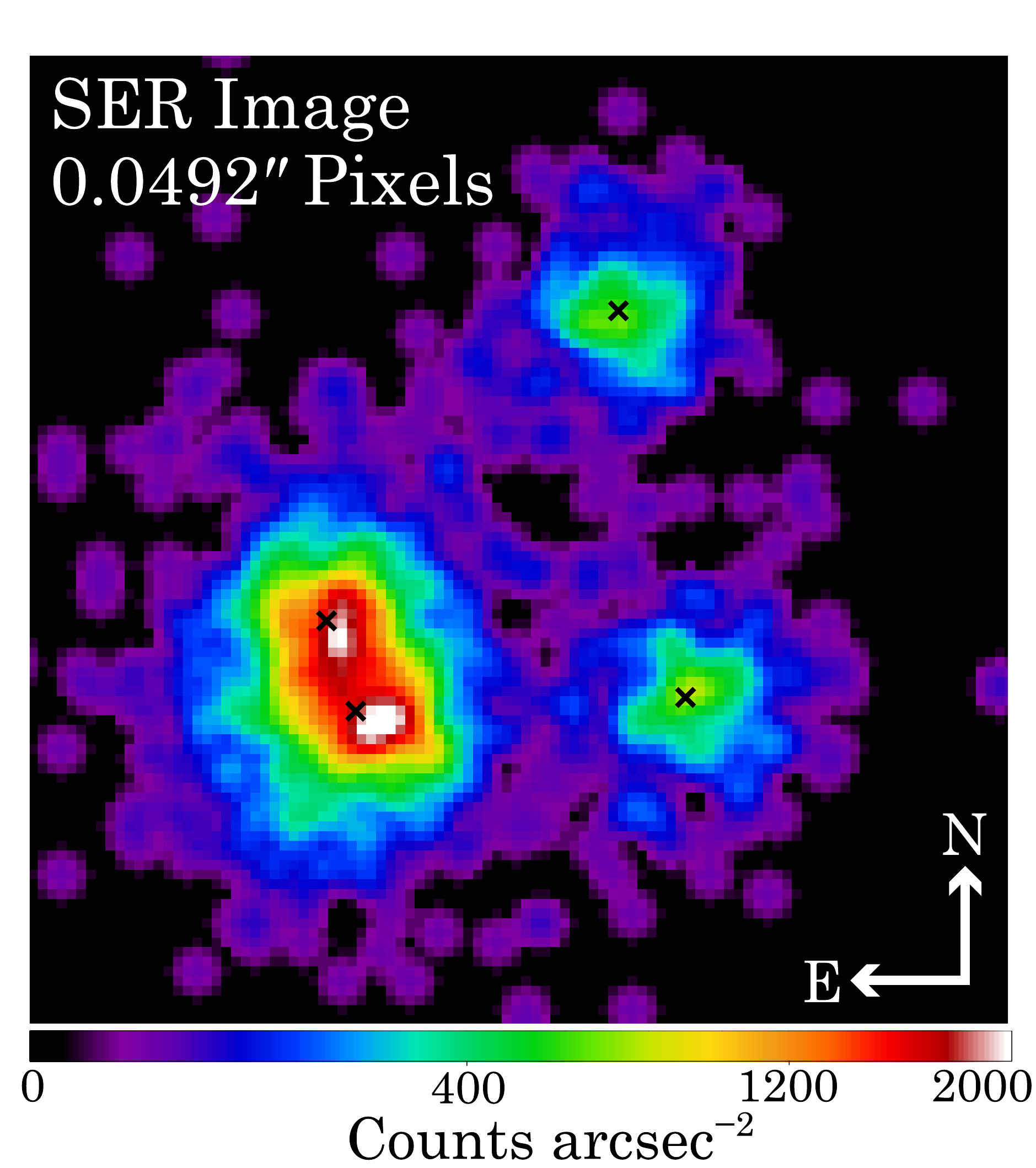}\hglue 0in\includegraphics[width=0.25\textwidth]{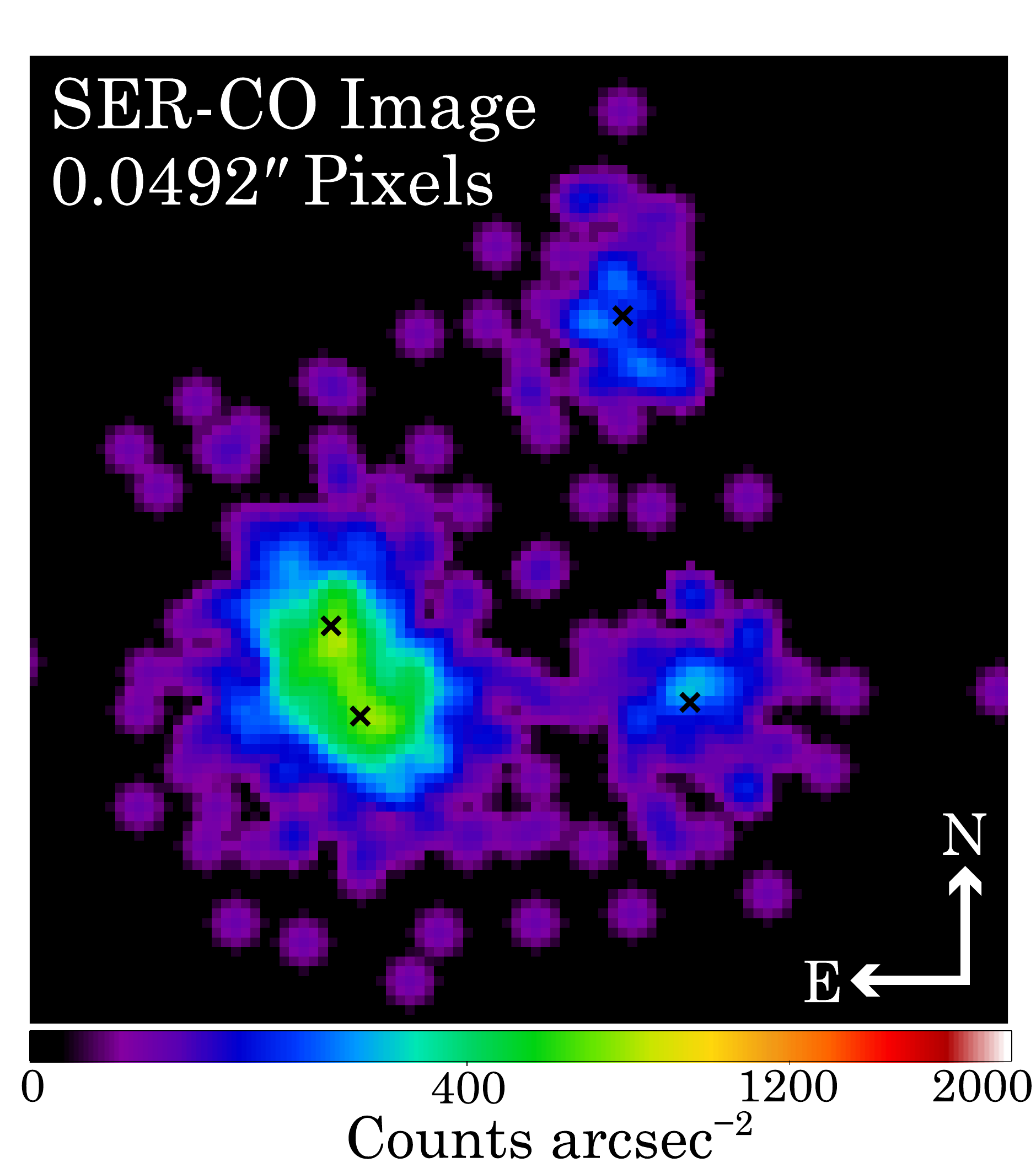}
\caption{Images of \pglens\ ObsID 7757 in the 0.3--8 keV band. The left image is 10\arcsec\ on a side and is made at the default resolution of 0\farcs492, which matches the physical pixel size of the ACIS detector.  The white box shows the region of the other three panels, which are 5\arcsec\ on a side and made at 10 times finer resolution than the default.  These images have been smoothed by a Gaussian with a 3-pixel FWHM for display purposes only.  The color maps are different for the large and small images, but in all cases the intensity scales as the square root of the surface brightness from 0 to 2000 counts arcsecond$^{-2}$. Note that the close pair is severely blended in the standard image but has two distinct peaks in the SER and SER-CO images.  The loss of counts in the SER-CO algorithm is evident.  Black crosses mark the best-fit locations of the $\beta$c model to the SER image.  See the text for details. }
\label{fig:serimgs}
\end{figure*}

\begin{figure}[h]
\centering
\includegraphics[width=0.47\textwidth]{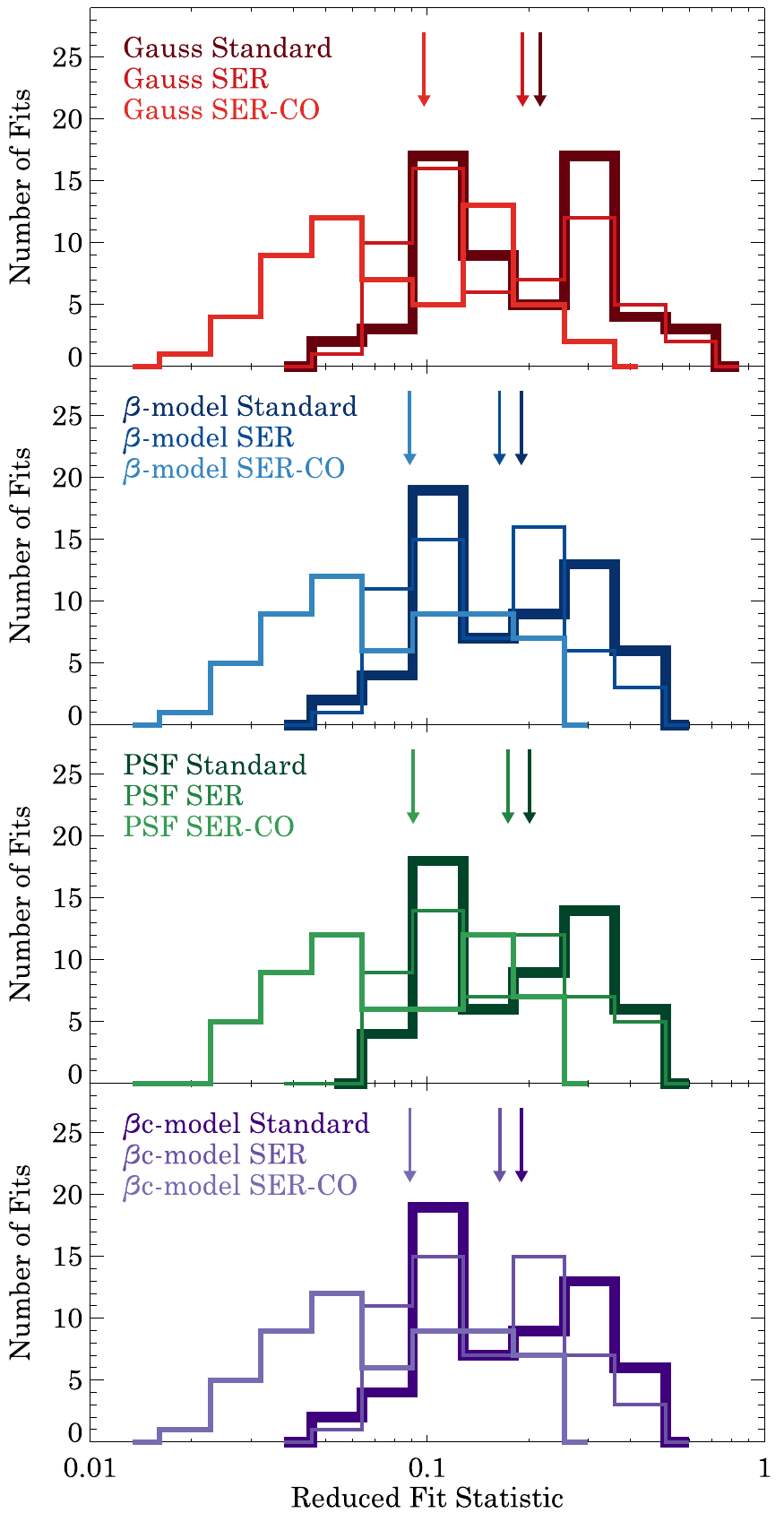}
\caption{Histograms of the reduced fit statistic for (top to bottom) Gaussian model fits, $\beta$-model fits, PSF fits, and $\beta$c-model fits to each of the images (standard, SER, and SER-CO).  The arrows indicate the mean value of the reduced statistic.  For the standard, SER, and SER-CO images, respectively, these are 0.21, 0.19, and 0.10 for the Gaussian fits, 0.19, 0.16, and 0.09 for the $\beta$-model fits, 0.20, 0.17, and 0.09 for the PSF fits, and 0.19, 0.16, and 0.09 for the $\beta$c-model fits.}
\label{fig:rstat}
\end{figure}

\subsection{Comparison of Standard, SER, and SER-CO Results}
\label{sec:sercomp}
As mentioned above, the aim of the SER algorithms is to improve the spatial resolution of the X-ray image.  They do this by using only split-pixel events so there is necessarily a loss of signal, and this needs to be weighed against the improved resolution.  To quantify the signal loss, we measure the number of 0.3--8 keV counts in the standard image, SER image and SER-CO image in a 6\farcs3 square region around the lensed quasars.  The distributions of the fraction of counts in the SER and SER-CO images compared to the standard image are shown in Figure~\ref{fig:sercounts}.  The SER algorithm is able to utilize about 75\% of the events on average, while the SER-CO image can utilize only about 25\%. 

The advantage of the SER algorithms is demonstrated in Figure~\ref{fig:serimgs}, which shows a standard image of \pglens\ made at the default resolution along with the standard, SER, and SER-CO images made at 0\farcs0492 pixel$^{-1}$. The blended close pair in the standard image is separated into two distinct peaks in the SER image.  

This improvement in resolution can be quantified by the width parameters of the Gaussian and $\beta$ models. Using the values from all observations of all systems, we find that the best-fit Gaussian FWHM is about 0\farcs06 smaller on average in the SER images compared to the standard images.  It is only about another 0\farcs02 smaller on average in the SER-CO images.  The best-fit $\beta$-model $r_0$ is about 0\farcs12 smaller on average in the SER images compared to the standard images, but also only about another 0\farcs02 smaller on average in the SER-CO images.  The results are nearly identical for the $\beta$c model: 0\farcs12 smaller in the SER images and only another 0\farcs02 smaller in the SER-CO images.  This small gain in resolution of $\sim$0\farcs02 of the SER-CO images comes at a large price in signal loss.

We compared the best-fit amplitudes of a specific model in the standard, SER, and SER-CO images and found reasonably good agreement among all four models.  There tended to be more outliers in the SER-CO image fits, likely due to decreased signal, but most amplitudes agreed within 1$\sigma$ uncertainties.   This good agreement is reassuring, but the real aim of exploring the SER and SER-CO images is the possibility of better constraining the amplitudes, i.e., reducing the uncertainty in the best-fit model amplitudes.  

In all cases, the model fits to the SER-CO have larger amplitude errors on average, again most likely due to the large reduction in signal inherent in using that algorithm.  In the Gaussian, $\beta$-, and $\beta$c-model fits, the amplitude errors in the SER fits and the standard fits are comparable (13\% for SER versus 12\% for standard in the Gaussian fits and 15\% versus 13\% in both the $\beta$- and $\beta$c-model fits), whereas the amplitude errors in the SER fits are somewhat smaller than those in the standard fits with the PSF model (12\% versus 17\%).

\begin{figure}
\centering
\includegraphics[width=0.47\textwidth]{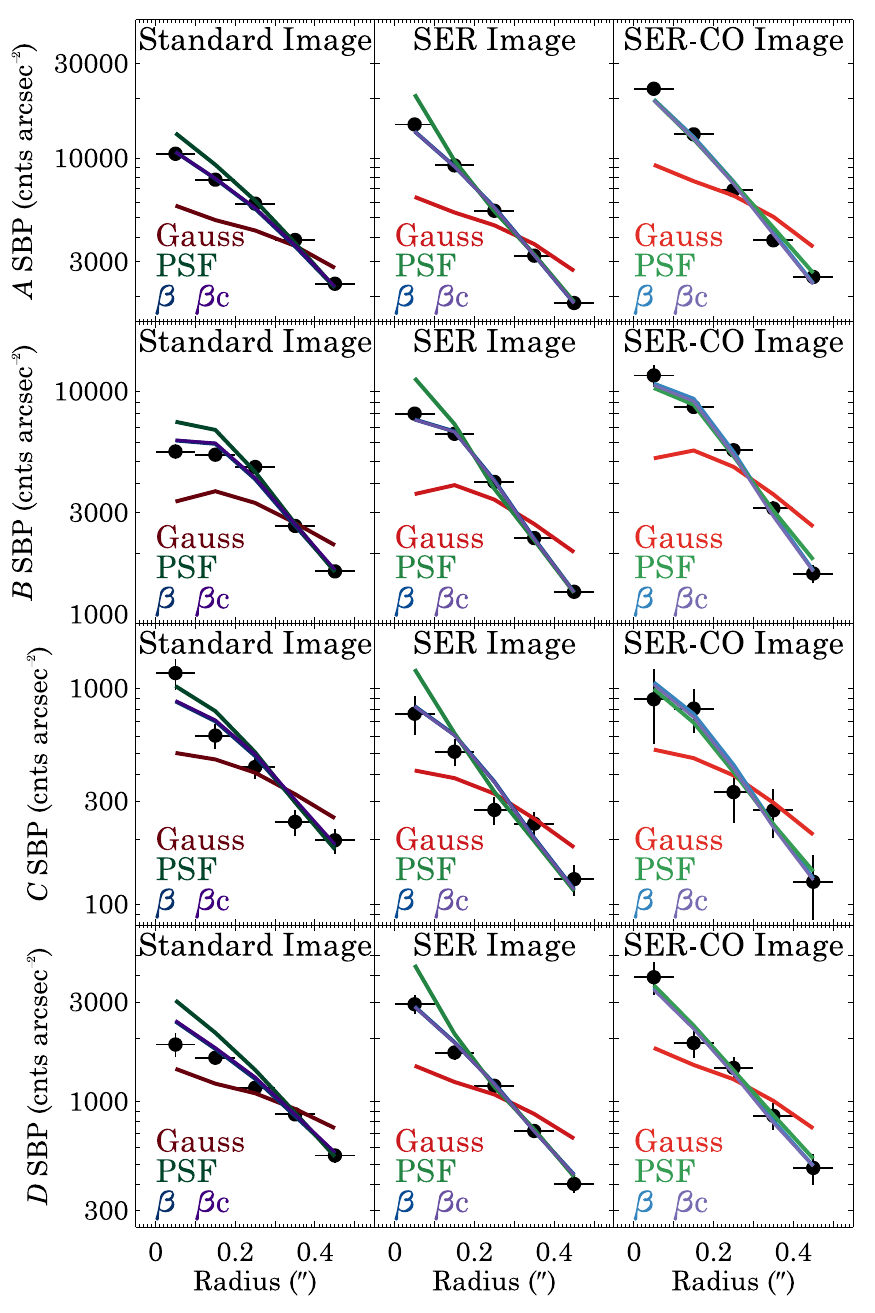}
\caption{Radial surface brightness profiles (SBPs) of the four images (top to bottom: $A$, $B$, $C$, and $D$) of \rxelevenlens\ from ObsID 9181.  Profiles are made from the (left to right) standard, SER, and SER-CO sky images.  The data and uncertainties are shown as black points with the horizontal bar indicating the bin width.  Profiles of the four model fits of each image are overlaid.  In most cases, the $\beta$- and $\beta$c-model profiles are indistinguishable.  Note that the $y$-axis is logarithmic.}
\label{fig:rprofs}
\end{figure}

\subsection{Comparison of Gaussian, $\beta$, Ray-traced PSF, and $\beta$c Models}
The most important feature of a model is how well it represents the data.  To explore this for the four models, we show histograms of the reduced ``cstat'' statistic in Figure~\ref{fig:rstat}.  The mean values are almost identical for all models, but the Gaussian ones tend to have tails to higher values of the reduced statistic than the others.

Another way to visualize the goodness of the image fit is by comparing the radial profiles of the data with each of the models.  To illustrate this, we choose the observation that has the highest number of counts, which is ObsID 9181 of \rxelevenlens.  The radial profile of each quasar image from the center to 0\farcs5 is shown in Figure~\ref{fig:rprofs}.  Overlaid are the radial profiles from the best-fit models of each type.  The Gaussian models are universally too squat, and the PSF models are often a poor fit in the center.  The $\beta$- and $\beta$c-model profiles do the best job of matching the data at all radii.  Note that these are not fits to the radial profiles; rather, they are radial profiles of the best-fit models overlaid on radial profiles of the data.

\subsection{Discussion of Fits and Choice of Best Image/Model Combination}
Considering the fit statistics and radial profiles, the $\beta$- and $\beta$c-models appear to be the best choices to represent the data.  Given the tight correlation seen in Figure~\ref{fig:r0alpha}, it is not surprising that both models give nearly identical results.  We have a slight preference for the $\beta$c-models because of the reduction in free parameters.  

Although the SER-CO fits have smaller reduced statistics on average, they have larger uncertainties and more outliers and are therefore not the ideal choice.  Between the standard and SER images, the uncertainties are comparable, as are the reduced statistics.  We favor the SER images for the task at hand because we believe that the increase in effective spatial resolution will provide higher fidelity results for the systems where the separation between quasar images is far less than 1\arcsec.

The results from the $\beta$c-model fits to the SER images are given in Table~\ref{tab:xrayfluxes}.  Each amplitude and uncertainty is reported as a fraction of the total, defined as the sum of the four amplitudes.  In many cases the amplitude errors are asymmetric.

\section{Dark Matter Determinations}
\label{sec:dm}
 Our determination of the fraction of stellar matter that makes up the total surface mass density for these systems relies on the analysis of microlensing magnification maps and follows the Bayesian methods of \citet{2009ApJ...697.1892P}.  Our specific method is worked out and discussed below.

\subsection{Microlensing Magnification Maps}
\label{sec:magmaps}
The four images of each quasar are either saddle-points or minima of the light travel time surface.  We denote the higher magnification minimum as the ``HM'' image and the lower magnification minimum as the ``LM'' image.  Likewise for the saddle point images, the higher magnification saddle point is  ``HS'', and the lower magnification saddle point is  ``LS.''  We have previously modeled all of these lens systems \citep{2007ApJ...661...19P,2011ApJ...729...34B} to determine the local convergence $\kappa$ and shear $\gamma$ for each of these images, which also gives the ``macrolensing'' magnification of each image. These parameters given in Table~\ref{tab:macromag} are provided by the models presented in \citep{2011ApJ...729...34B}.
\begin{deluxetable*}{llclllclllclllclll}[t!]
\tablewidth{0pt}
\tablecaption{Lensing Galaxy Parameters}
\tablehead{\multicolumn{2}{c}{} &  \multicolumn{4}{c}{HM} & \multicolumn{4}{c}{HS} &\multicolumn{4}{c}{LM} & \multicolumn{4}{c}{LS}\\
\colhead{System} & \colhead{$z_l$} & \colhead{Im.} & \colhead{$\kappa$} & \colhead{$\gamma$} & \colhead{Magnif.}& \colhead{Im.} & \colhead{$\kappa$} & \colhead{$\gamma$} & \colhead{Magnif.} & \colhead{Im.} & \colhead{$\kappa$} & \colhead{$\gamma$} & \colhead{Magnif.} & \colhead{Im.} & \colhead{$\kappa$} & \colhead{$\gamma$} & \colhead{Magnif.}} 
\startdata
          \hetwolens & 0.52 & $  A$ & 0.472 & 0.416 & $+9.46$ & $  B$ & 0.510 & 0.587 & $-9.57$ & $  C$ & 0.440 & 0.334 & $+4.95$ & $  D$ & 1.070 & 0.864 & $-1.35$ \\
             \mglens & 0.96 & $A_1$ & 0.481 & 0.475 & $+22.9$ & $A_2$ & 0.496 & 0.544 & $-23.9$ & $  B$ & 0.478 & 0.335 & $+6.24$ & $  C$ & 0.618 & 0.684 & $-3.11$ \\
         \hefourlens & 0.46 & $  C$ & 0.463 & 0.394 & $+7.51$ & $  B$ & 0.520 & 0.598 & $-7.86$ & $  A$ & 0.460 & 0.390 & $+7.17$ & $  D$ & 0.559 & 0.637 & $-4.73$ \\
         \rxninelens & 0.77 & $  B$ & 0.575 & 0.299 & $+11.0$ & $  A$ & 0.633 & 0.550 & $-5.96$ & $  D$ & 0.286 & 0.055 & $+1.97$ & $  C$ & 0.650 & 0.568 & $-5.00$ \\
       \sdssninelens & 0.39 & $  A$ & 0.490 & 0.440 & $+15.0$ & $  D$ & 0.517 & 0.557 & $-13.0$ & $  B$ & 0.450 & 0.390 & $+6.65$ & $  C$ & 0.546 & 0.599 & $-6.55$ \\
       \heelevenlens & 0.6$^\dag$ & $  B$ & 0.484 & 0.450 & $+15.7$ & $  D$ & 0.510 & 0.548 & $-16.6$ & $  A$ & 0.477 & 0.441 & $+12.6$ & $  C$ & 0.531 & 0.570 & $-9.53$ \\
             \pglens & 0.31 & $A_1$ & 0.537 & 0.405 & $+19.9$ & $A_2$ & 0.556 & 0.500 & $-18.9$ & $  C$ & 0.472 & 0.287 & $+5.09$ & $  B$ & 0.658 & 0.643 & $-3.37$ \\
       \rxelevenlens & 0.30 & $  B$ & 0.423 & 0.507 & $+13.2$ & $  A$ & 0.442 & 0.597 & $-22.2$ & $  C$ & 0.422 & 0.504 & $+12.5$ & $  D$ & 0.834 & 0.989 & $-1.05$ \\
     \sdsselevenlens & 0.45 & $  A$ & 0.465 & 0.384 & $+7.21$ & $  D$ & 0.523 & 0.614 & $-6.69$ & $  C$ & 0.438 & 0.349 & $+5.15$ & $  B$ & 0.578 & 0.673 & $-3.64$ \\
              \hlens & 0.8$^\dag$ & $  B$ & 0.454 & 0.359 & $+5.91$ & $  A$ & 0.531 & 0.634 & $-5.49$ & $  C$ & 0.441 & 0.343 & $+5.13$ & $  D$ & 0.576 & 0.680 & $-3.54$ \\
              \blens & 0.34 & $  A$ & 0.371 & 0.532 & $+8.88$ & $  B$ & 0.400 & 0.666 & $-12.0$ & $  C$ & 0.360 & 0.485 & $+5.73$ & $  D$ & 1.530 & 1.800 & $-0.34$ \\
   \wfitwentysixlens & 0.4$^\dag$ & $A_1$ & 0.499 & 0.422 & $+13.7$ & $A_2$ & 0.528 & 0.557 & $-11.4$ & $  B$ & 0.405 & 0.299 & $+3.78$ & $  C$ & 0.579 & 0.653 & $-4.01$ \\
 \wfithirtythreelens & 0.66 & $A_1$ & 0.513 & 0.267 & $+6.03$ & $A_2$ & 0.621 & 0.638 & $-3.80$ & $  B$ & 0.416 & 0.290 & $+3.89$ & $  C$ & 0.650 & 0.727 & $-2.46$ \\
              \qlens & 0.04 & $  A$ & 0.400 & 0.400 & $+5.00$ & $  D$ & 0.617 & 0.617 & $-4.27$ & $  B$ & 0.385 & 0.385 & $+4.35$ & $  C$ & 0.721 & 0.721 & $-2.26$
\enddata
\tablerefs{\citet{2011ApJ...729...34B}.}
\tablecomments{$^\dag$ Estimated. See the text for details.}
\label{tab:macromag}
\end{deluxetable*}

These large-scale lens models can give only the total $\kappa$ at the site of each image without regard to the form of the matter present.  We generate a series of 12 custom microlensing maps for each image by assuming that some fraction of $\kappa$ is in a clumpy component (stars) and the rest is in a smooth component (dark matter).  We use a logarithmic sequence of stellar fractions ($S_j$): 1.47\%, 2.15\%, 3.16\%, 4.64\%, 6.81\%, 10\%, 14.68\%, 21.5\%, 31.62\%, 46.4\%, 68.13\%, and 100\%. 

In total, 672 microlensing maps were produced using the ``microlens'' ray-tracing code \citep{1990PhDT.......180W, 1990ApJ...352..407W, 1999JCoAM.109..353W}.  These magnification maps are constructed in the source plane, and their centers are referenced to the location of one of the quasar images. They show the effects of microlensing magnification (due to the sum of all the microimages) for a source location anywhere within the map.  The mean macrolensing magnification, due to the smooth lensing potential, has been subtracted off.  Each map is 2000 $\times$ 2000 pixels, with an outer scale of 20 \rein\ and a pixel size of 0.01 \rein, where \rein\ is the Einstein radius of a microlensing star of average mass.  The stars are drawn from a mass function similar to the well-known one of \citet{2001MNRAS.322..231K}.  The mass function runs from 0.08\Msun\ to 1.5\Msun\ with a break at 0.5\Msun\ and logarithmic slopes of $-1.8$ and $-2.7$ below and above the break, respectively.  The average mass of a microlensing star is 0.247\Msun, and the stellar mass above and below which 50\% of the mass lies is 0.335\Msun.

\begin{figure*}
\centering
\includegraphics[width=\textwidth]{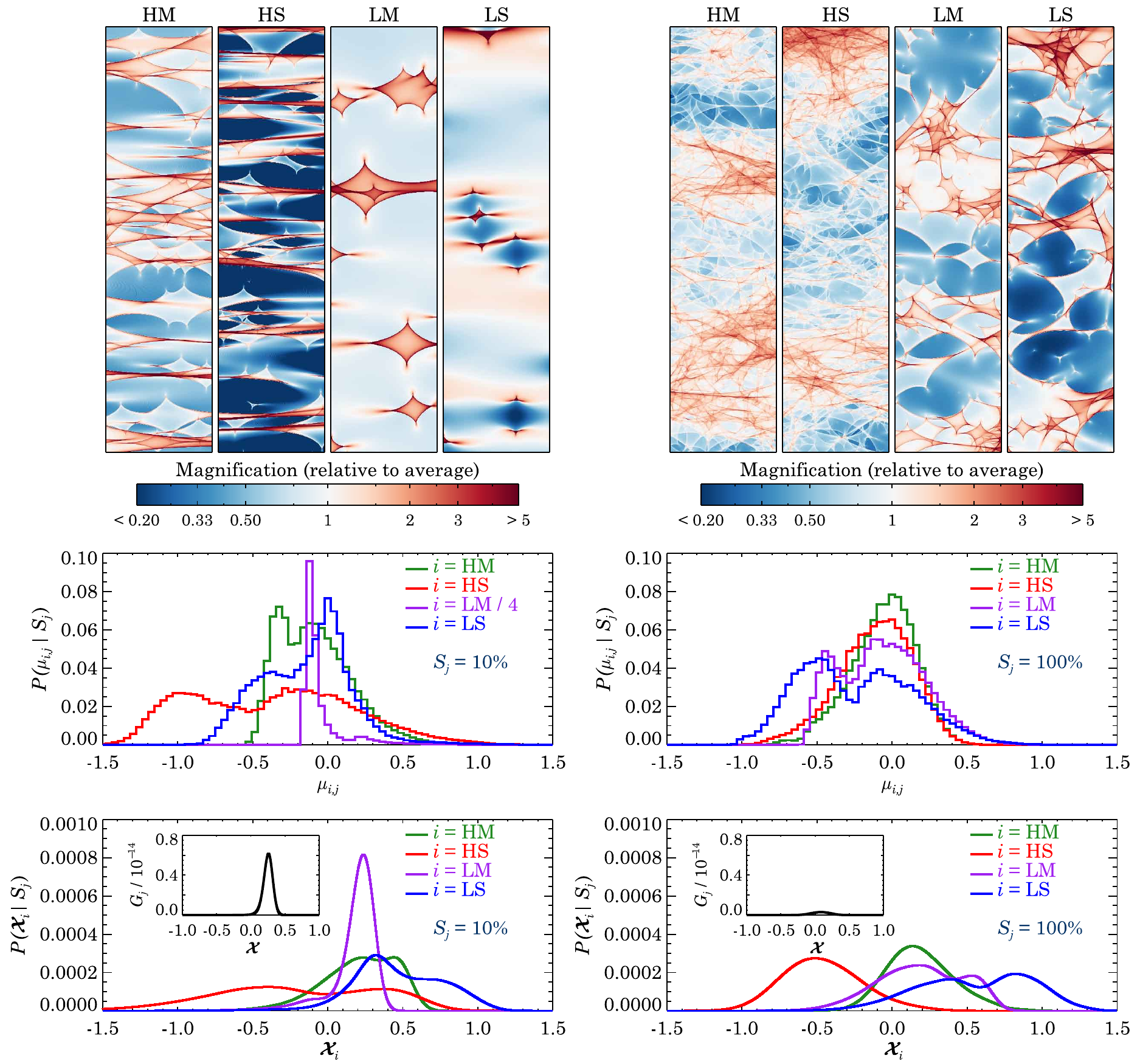}
\caption{{Top:} small portions (\sfrac{1}{16}) of the full microlensing magnification maps for each of the four images of \pglens\ for both $S_j = 10\%$ stars (left) and $S_j = 100\%$ stars (right).  These $250\times1000$ pixel segments illustrate the microlensing differences due to image type and stellar fraction. {Middle:} normalized histograms of the logarithm of the pixel values ($\mu_{i,j}$) in each microlensing magnification map.  {Bottom}: convolution of those histograms with the probability functions of the X-ray flux of each image using the data from ObsID 363.  These give the independent probability distributions for the intrinsic flux of the quasar, $\lfx = \log_{10}(F_\mathrm{X,intr} / 10^{-14}~\ergcms)$.  Plotted in the inset is their product $G_j$.  See the text for details.}
\label{fig:maps}
\end{figure*}

Figure~\ref{fig:maps} shows portions of each of the four microlensing maps (HM, HS, LM, and LS) produced for \pglens\ for stellar fractions of  both 10\% and 100\% to illustrate the differences among the microlensing maps.  For each map, a histogram of the logarithm of the magnification values is made and normalized, and this is used as the probability distribution for microlensing effects $P(\mu_{i,j}|S_j)$ where $\mu_{i,j} = \log_{10}(\mathrm{micromag}_{i,j})$,  $i\in\{\mathrm{HM},\mathrm{HS},\mathrm{LM},\mathrm{LS}\}$, and $S_j$ is one of the stellar fractions listed above.  For convenience, we also define $m_{i,j} = -\mu_{i,j}$.  These normalized histograms are shown in the bottom panels of  Figure~\ref{fig:maps}.

\subsection{Bayesian Analysis}
Our goal is to determine the probability of each stellar fraction $S_j$ for a lensing galaxy.   Our measurements of the X-ray fluxes of the four images divided by their respective macrolensing magnifications give four estimates of the intrinsic flux $F_\mathrm{X,intr}$ of the quasar.  We use conditional probability to express $P(S_j)$ as
\begin{equation}
P(S_j) = \sum_\lfx P(S_j | \Xhm,\Xhs,\Xlm,\Xls) P(\lfx)
\label{eq:cond}
\end{equation}
where $\lfx = \log_{10}(F_\mathrm{X,intr} / F_\mathrm{norm})$ and $\lfx_i$ indicates the estimate of \lfx\ from image $i$.  We choose  $F_\mathrm{norm} = 10^{-14}\ \ergcms$, which has no effect on the analysis.

We use Bayes's theorem to express 
\begin{multline} 
P(S_j | \Xhm,\Xhs,\Xlm,\Xls) =  \\
\frac{P( \Xhm,\Xhs,\Xlm,\Xls|S_j)P_\mathrm{pr}(S_j)}{\sum_j P( \Xhm,\Xhs,\Xlm,\Xls|S_j)P_\mathrm{pr}(S_j)}
\label{eq:bayes}
\end{multline}
where $P_\mathrm{pr}(S_j)$ is the a priori probability of $S_j$ and the denominator is a normalization term.  We take  $P_\mathrm{pr}(S_j)$ to be uniform and combine it with the denominator as the constant $A$ in what follows.  We compute it by ensuring that $\sum_j P(S_j) = 1$.

\begin{figure*}
\centering
\includegraphics[width=\textwidth]{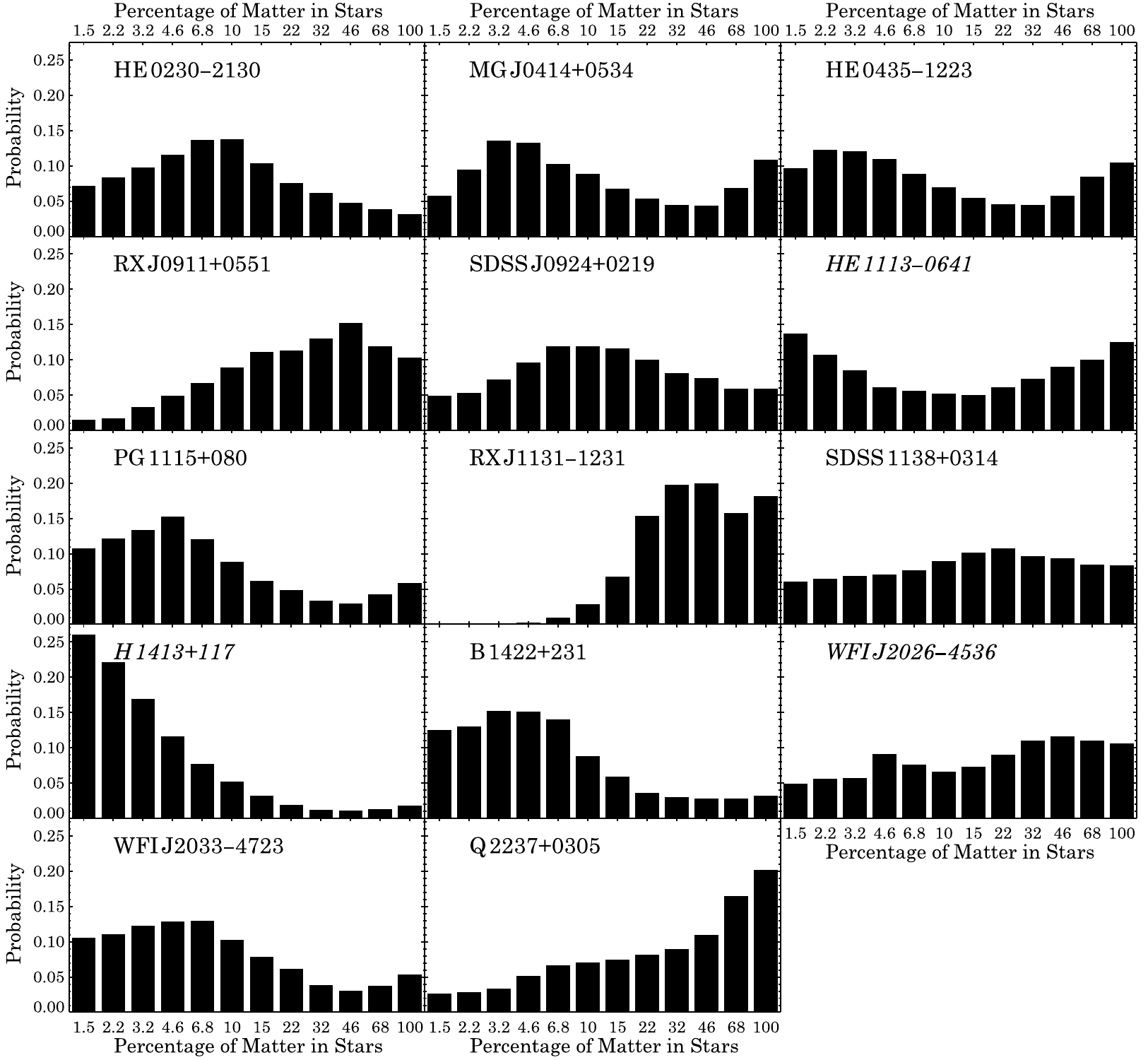}
\caption{Probability distributions for the stellar fraction, $S_j$, at the characteristic radial distance $R_c$ from the center of the lensing galaxy for 14 quadruply lensed quasar systems.  Those labeled in {\it italics} do not have a measured lens redshift $z_l$.}
\label{fig:indivprobs}
\end{figure*}

Because the four $\lfx_i$ are physically distinct, their probabilities are independent from each other, and we can express
\begin{equation}
P( \Xhm,\Xhs,\Xlm,\Xls|S_j) = \prod_i P(\lfx_i|S_j)~~~~.
\label{eq:indep}
\end{equation}
Substituting Equations~(\ref{eq:bayes}) and (\ref{eq:indep}) into Equation~(\ref{eq:cond}), we arrive at
\begin{equation}
P(S_j) = A \sum_\lfx \prod_i P(\lfx_i|S_j) P(\lfx)
\label{eq:psj}
\end{equation}
and what remains is to calculate  $P(\lfx_i|S_j)$ and $P(\lfx)$.

The probability of the intrinsic flux of a quasar $P(\lfx)$ can be determined from the number counts obtained from deep studies of the X-ray background.  We use the results from \citet{2001ApJ...551..624G} that $N(>\Fx) \sim \Fx^{-0.85}$, but we note this has little impact on the analysis.  A uniform distribution would produce nearly identical results.

We estimate the intrinsic flux $F_\mathrm{X,intr}$ from the measured flux ($f_{\mathrm{X},i})$ of an image and the lensing effects, both macrolensing and microlensing.  The measured flux is
\begin{equation}
f_{\mathrm{X},i} = F_\mathrm{X,intr} \times \mathcal{M}_i \times 10^{\mu_{i,j}}
\end{equation}
where $\mathcal{M}_i$ is the macro-magnification of image $i$.  We define 
\begin{equation}
x_i = \log_{10}([f_{\mathrm{X},i}/F_\mathrm{norm}]/\mathcal{M}_i)
\end{equation}
which allows us to write
\begin{equation}
\lfx_i = x_i + m_{i,j}~~~~~.
\end{equation}
Because the probability of the sum of two random variables is the convolution of their individual probabilities, we can express
\begin{equation} \label{eq:convol}
\begin{split}
P(\lfx_i|S_j) & =  P(x_i + m_{i,j} | S_j)\\
                      & = P(x_i) * P(m_{i,j}|S_j)
\end{split}
\end{equation}
where $P(x_i)$ comes from the uncertainties on the flux measurements of the images and the $P(m_{i,j}|S_j)$ are the reverse of $P(\mu_{i,j}|S_j)$, the normalized histograms of the microlensing maps, as discussed above.

We do not measure the $f_{\mathrm{X},i}$ directly, though; rather, we obtain them by multiplying the total flux of all four images (via spectral fitting) and the individual fractions of the total (via two-dimensional image fitting).  We define 
\begin{equation}
T = \log_{10}(F_\mathrm{X,tot}/F_\mathrm{norm})
\end{equation} 
and 
\begin{equation}
r_i = \log_{10}(\mathrm{frac}_i / \mathcal{M}_i)
\end{equation}
so that 
\begin{equation}
x_i = r_i + T~~~~~.
\end{equation}
Again, using the property of the sum of two random variables, we express
\begin{equation} \label{eq:convolx}
\begin{split}
P(x_i) & =  P(r_i + T)\\
            & = P(r_i) * P(T)
\end{split}
\end{equation}
where the probability distributions $r_i$ are asymmetric Gaussians with standard deviations equal to the 1$\sigma$ uncertainties in the image fractions (Table~\ref{tab:xrayfluxes}) and $T$ is a symmetric Gaussian with a standard deviation equal to the uncertainty in $f_\mathrm{X,tot}$ (Table~\ref{tab:xrayfluxes}).  Introducing notation $G_j$ and using Equations~(\ref{eq:convol}) and (\ref{eq:convolx}), we have
\begin{equation} \label{eq:plfx}
\begin{split}
G_j & =\prod_i P(\lfx_i|S_j)\\
             & = \prod_i P(r_i) * P(T) * P(m_{i,j}|S_j)
\end{split}
\end{equation}
which can be seen in the bottom panels of Figure~\ref{fig:maps} using values from ObsID 363.  

All of the above has been worked out for a single observation of a system, but several systems have been observed multiple times with \chandra.  We combine these multiple observations using conditional probability:
\begin{equation}
P(S_j) = \sum_k P(S_j|\mathrm{obs}_k) P(\mathrm{obs}_k)
\label{eq:condobs}
\end{equation}
where we take $P(\mathrm{obs}_k)$ as a weighting factor (normalized to unity) that combines two measures of the effectiveness of the observation to provide unique and useful information.  

The first ingredient in $P(\mathrm{obs}_k)$ concerns the uniqueness of the information from the observation.  Over time, the proper motions of the lensing galaxy and background quasar, as well as the internal motions of the microlensing stars, can be thought of as an effective motion of the source through the field of the microlensing map \citep{2000MNRAS.312..843W}.  The more time between observations, the higher the chance that the source is in a different enough region of the map to be considered an independent sampling of it.  We therefore include a term in $P(\mathrm{obs}_k)$ proportional to how isolated in time the observation is, defined as the sum of the intervals between the observation and all other observations.

The second ingredient in $P(\mathrm{obs}_k)$ is based on the quality of the information that the observation provides.  Observations which yield tight constraints on the individual fractions and the total flux consequently give much better defined probability functions for the stellar fraction (we point out specific examples below).  We use the measured uncertainties (Table~\ref{tab:xrayfluxes}) on the fractions (symmetrized) and the total flux to calculate this.  Our full expression is
\begin{equation}
P(\mathrm{obs}_k) = B\left(\sum_{l\neq k} \left|t_k - t_l\right| \right) \ \prod_i \frac{\mathrm{frac}_{i,k}}{\sigma_{\mathrm{frac}_{i,k}}}\ \frac{F_{\mathrm{X,tot},k}}{\sigma_{F_{\mathrm{X,tot},k}}}
\end{equation}
where $t_k$ is the epoch of observation $k$ and $B$ is a normalization constant such that $\sum_k P(\mathrm{obs}_k) = 1$.

Using Equations~(\ref{eq:convol}), (\ref{eq:convolx}) and (\ref{eq:condobs}), we can express Equation~(\ref{eq:psj}) in terms of observables, the microlensing magnification map histograms, and the intrinsic flux probability from the deep field quasar number counts:
\begin{equation}
P(S_j) = \sum_k  A_k \left(\sum_\lfx G_{j,k} P(\lfx)\right) P(\mathrm{obs}_k)~~~.
\label{eq:finalprob}
\end{equation}

These probabilities are plotted for each of the 14 lensing galaxies in Figure~\ref{fig:indivprobs} as functions of the stellar mass fraction.  

The effect of poorly constrained  image fluxes is easily seen in the nearly flat probability distribution of \sdsselevenlens, which has only one \chandra\ observation, in which the average uncertainty of the image fractions is $\sim$70\%. Compare this to \hetwolens, which also has only one \chandra\ observation, but in which the average image fraction uncertainty is $\sim$20\%.

\subsection{Combined Analysis}
\label{sec:combinedanalysis}
We would like to consider each lensing galaxy as a typical member of an ensemble, each with roughly the same configuration such that we are probing the matter content at roughly the same radial distance $R$ from the center of the lensing galaxy.   To calculate these distances in physical units, we use the angular measurements of the images and galaxies available on the CASTLES Web site\footnote{\url{http://www.cfa.harvard.edu/castles/}} along with the redshifts to the lensing galaxies, $z_l$.

\begin{figure}
\centering
\includegraphics[width=0.47\textwidth]{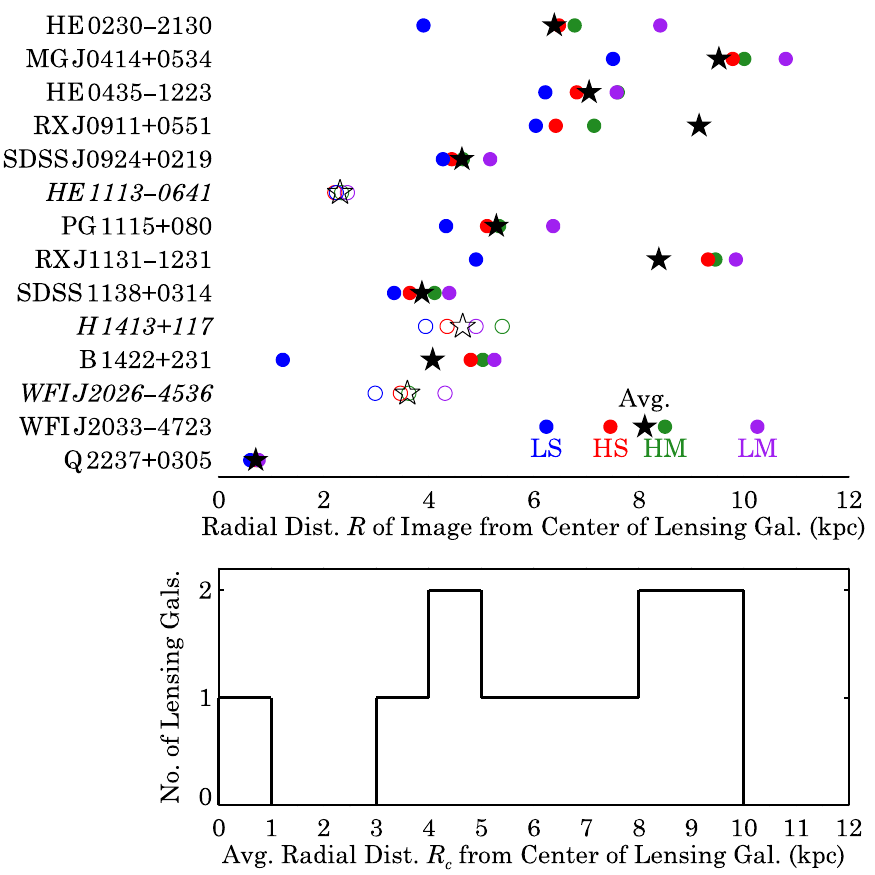}
\caption{{Top:} distances of quasar images (circles) and their mean (stars) from center of lensing galaxy. The redshifts of the lensing galaxies of \heelevenlens, \hlens, and \wfitwentysixlens\ have not been measured and were taken to be 0.7, 0.8, and 0.4, respectively.  Their symbols are shown in outline.  The LM image of \rxninelens\ is at a radial distance of 17 kpc and is not shown. {Bottom:} distribution of mean radial distance of images, $R_c$, for the 11 lensing galaxies with known redshift.}
\label{fig:impact}
\end{figure}

We take the arithmetic mean of the four impact parameters where the images form in the lensing galaxy as a characteristic radial distance, $R_c$.   Most of the systems have $R_c$ within a factor of a few of each other except for \qlens, in which the images form at a mean $R_c$ of 0.7 kpc, about an order of magnitude less than the mean $R_c$ of 6.6 kpc of the other systems (see Figure~\ref{fig:impact}).  We exclude \qlens\ from the rest of the analysis.  We note that, had we used the geometric means instead, the numbers would be very similar.  The images in \qlens\ form at a geometric mean radial distance of 0.7 kpc, and the geometric mean of the radial distances for the rest of the ensemble is 6.1 kpc.

Unfortunately, $z_l$ is not known for \heelevenlens, \hlens, and \wfitwentysixlens\  so $R$ cannot be calculated for the images of these systems.  There have been indications of lenses of \hlens\ at redshifts of 0.8, 1.4, and 1.7 \citep{1988Natur.334..325M,1998A&A...339L..65K, 2004A&A...428..741F}.  \citet{2004AJ....127.2617M} estimate a redshift of 0.4 for the lensing galaxy of \wfitwentysixlens, and \citet{2008AJ....135..374B} estimate $z_l = 0.7$ for \heelevenlens.  These values are all comparable to the redshifts of the other lensing galaxies, unlike \qlens\ with $z_l = 0.04$, and it is reasonable to assume that their impact parameters are also comparable.  We therefore include them in the joint analysis.  The radial distances of the images in these systems are shown with outlined symbols in the top of Figure~\ref{fig:impact} assuming $z_l$ of 0.7, 0.8, and 0.4 for \heelevenlens, \hlens, and \wfitwentysixlens, respectively; they are not included in the histogram in the bottom panel of Figure~\ref{fig:impact}.

\begin{figure}
\centering
\includegraphics[width=0.47\textwidth]{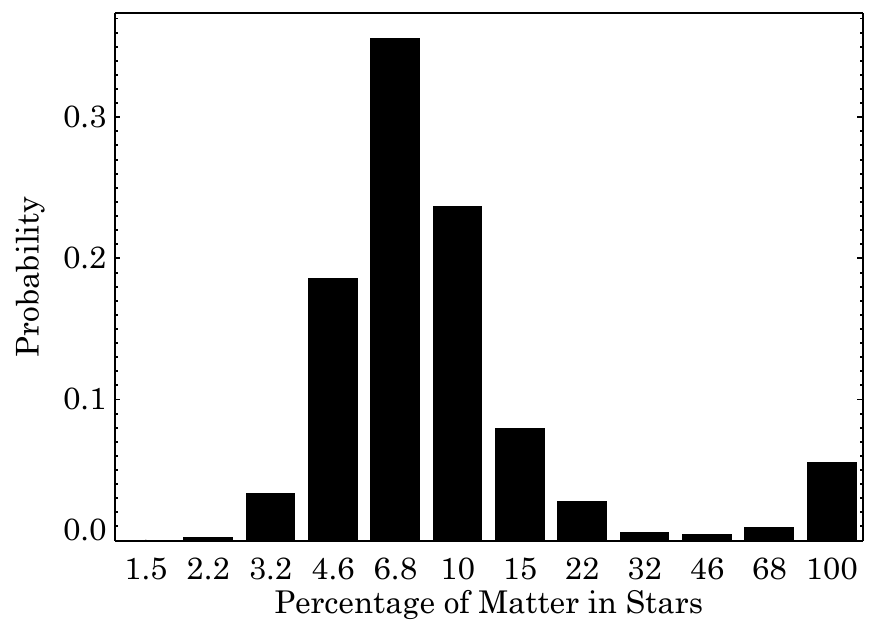}
\caption{Overall probability distribution for the percentage of matter in stars including all the X-ray observations for 13 quadruple lens systems (we do not include \qlens ---see Section \ref{sec:combinedanalysis}). The most likely value for the stellar contribution is 6.8\% at a mean impact parameter of 6.6 kpc.}
\label{fig:jointprob}
\end{figure}

\begin{figure}[]
\centering
\includegraphics[width=0.47\textwidth]{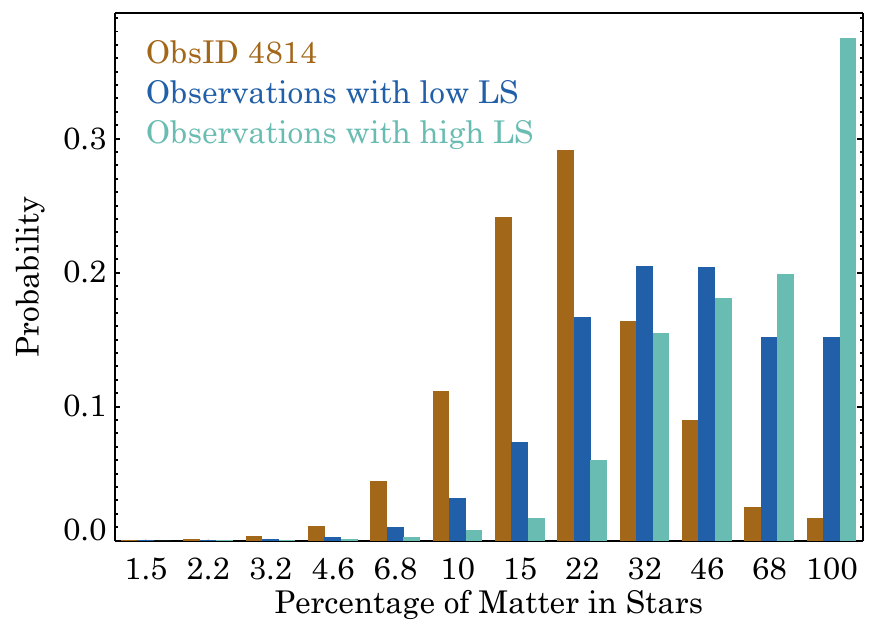}
\caption{Normalized probability distributions for the stellar fraction in the lensing galaxy of \rxelevenlens\ based on splitting the observations into three groups.  The first group, shown in brown, contains only the observation from 2004 (ObsID 4814).  The other two groups contain the observations from 2006--2008, split into whether the LS image fraction (given in Table~\ref{tab:xrayfluxes}) was higher (blue) or lower (teal) than 0.05.}
\label{fig:rxj1131}
\end{figure}

\begin{figure}[t!]
\centering
\includegraphics[width=0.47\textwidth]{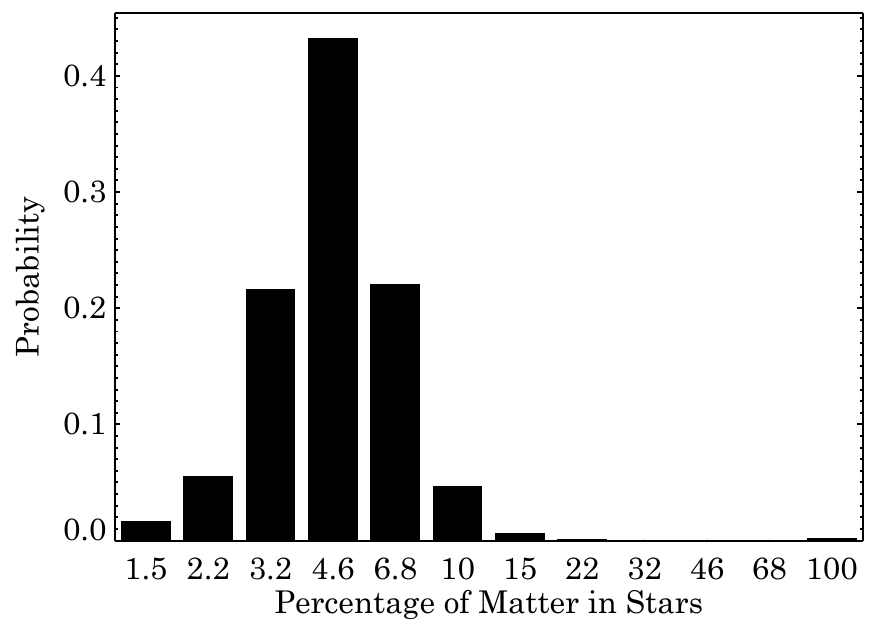}
\caption{As in Figure~\ref{fig:jointprob} but also excluding \rxelevenlens.}
\label{fig:probnorxj1131}
\end{figure}

For the 10 systems other than \qlens\ with known $z_l$, their mean impact parameters $R_c$ are within a factor of 2.5 of each other.  If we consider all individual $R$ in these 10 systems, the spread is nominally a factor of 14.  Excluding the two extrema (the LS image in \blens\ at $R=1.2$~kpc and the LM image in \rxninelens\ at $R=17$~kpc), the spread in $R$ among the 10 systems with known $z_l$ is only a factor of 3.2.  As we discuss in Section \ref{sec:dmvdist}, this is a small enough range in $R_c$ and $R$ that the ratio of stellar matter to dark matter is expected to vary by only 1.6 over this interval, and we feel comfortable combining the individual results to obtain an ensemble result for a mean $R_c$ of $\langle R_c \rangle  = 6.6~\mathrm{kpc}$.  

We form the joint probability function of the ensemble---excluding \qlens ---by multiplying together the individual probability functions of the 13 lensing galaxies (shown in Figure~\ref{fig:indivprobs}), and normalizing.  The results are displayed in Figure~\ref{fig:jointprob}, which shows the joint probability distribution of the ensemble for the percentage of matter in stars at a mean impact parameter of 6.6 kpc.   The highest peak of this discrete distribution occurs at 6.8\% stellar matter (93.2\% dark matter), and the interpolated peak occurs at $6.3\%\pm0.3\%$ stellar matter (93.7\% dark matter).

\section{Discussion}
\label{sec:discuss}
\subsection{\rxelevenlens}
\label{sec:rxj1131}
When exploring the results of the individual observations shown in Figure~\ref{fig:indivprobs}, we noticed that \rxelevenlens\ had one of the highest probabilities for a 100\% stellar fraction, after \qlens.  This is surprising given that the first \chandra\ observation of \rxelevenlens\ displayed a strong signature of significant dark matter presence: a highly suppressed saddle-point image \citep{2006ApJ...640..569B}.  We examined the \rxelevenlens\ probabilities on an observation by observation basis and found that, indeed, the first observation (ObsID 4814) strongly favored a stellar fraction of 22\%.  The other observations favored higher stellar fractions, either with a  roughly flat distribution above 22\% stars or a strong peak at 100\% stars.  

We noticed a correlation that those observations with a flat distribution above 22\% were the ones that had lower LS fractions, and the handful of observations (six) that peaked at 100\% stars were the ones with an LS fraction $>$0.05 in Table~\ref{tab:xrayfluxes}.  We separately analyzed these two groups, and the results are shown, along with ObsID 4814, in Figure~\ref{fig:rxj1131}.   

\rxelevenlens\ has the most X-ray observations and displays interesting behavior.  The HS image evolved from being strongly demagnified by microlensing to being strongly magnified by microlensing, and the LS image shows microlensing variations of over a factor of two.  It may be that our snapshot analysis is not appropriate for such complex behavior.  Our treatment of each observation separately and combination of their weighted results discards information on temporal evolution.  An analysis that assesses the probability of a certain stellar fraction, $S_j$, to produce the entirety of the observations, similar to the Bayesian III method in \citet{2009ApJ...697.1892P} may be more appropriate but is beyond the scope of this work.

Although we have some minor concerns about the robustness of the \rxelevenlens\ stellar fraction probabilities, their effect on our joint analysis shown in Figure~\ref{fig:jointprob} is minor.  If we perform the analysis without \rxelevenlens, we see that the most probable stellar fraction is slightly lower (the interpolated peak is at $4.6\%\pm0.2\%$), and the probability of 100\% stars is near zero (Figure~\ref{fig:probnorxj1131}).

\subsection{Dark Matter Fraction versus Radial Distance}
\label{sec:dmvdist}
Our measurements of the stellar mass fraction pertain to the impact parameter, $R$, that the quad images make with respect to the lensing galaxy, and span a range of $\sim$3--11 kpc.  For any given impact parameter, $R$, a large range of radial distances in three dimensions (i.e., for all $r > R$) is probed within the lensing galaxy.  Since the percentage of mass in stars is expected to decrease with increasing $r$, we would like to ascertain whether our ensemble average likelihood distribution for the dark matter fraction (see Figure~\ref{fig:jointprob}) is well defined, or whether we should expect to see a decreasing progression of star fraction with increasing mean impact parameter, $R_c$.

Following \citet{2009ApJ...703L..51K} and \citet{2010ApJ...708..750S}, we express the three-dimensional light density in an elliptical galaxy as
\begin{equation}
I(r) = I_{\rm S0} \left(\frac{r}{r_0}\right)^{-\delta}
\label{eq:stars}
\end{equation}
where $I_{\rm S0}$ and $r_0$ are constants for a given galaxy, and $\delta$ is a more nearly universal constant which \citet{2010ApJ...708..750S} determined to be
\begin{equation}
\delta = 2.4 \pm 0.11
\end{equation}
based on 54 lenses from the SLACS survey \citep[e.g.,][]{2006ApJ...640..662T, 2006EAS....20..161K, 2007ApJ...667..176G}.  The value of 0.11 is supposed to represent the rms variation in $\delta$ among different galaxies, rather than an uncertainty in the mean value of $\delta$. We assume that this power law holds over the radial interval $r \simeq$~1--10~kpc.  We also assume that the stellar mass function and evolutionary states of the stars are distance-independent, so that $I(r)$ also represents the stellar mass density.

Similarly, \citet{2009ApJ...703L..51K} and \citet{2010ApJ...708..750S} took the {\em total} mass density to be of the form:
\begin{equation}
\rho(r) = \rho_0\left(\frac{r}{r_0}\right)^{-\alpha}
\label{eq:rho}
\end{equation}
with constants that are analogous to those in Equation (4); $\alpha$ is found to be
\begin{equation}
\alpha = 1.96 \pm 0.08
\end{equation}
again based on the SLACS survey \citep[e.g.,][]{2006ApJ...640..662T, 2006EAS....20..161K, 2007ApJ...667..176G, 2009ApJ...703L..51K, 2010ApJ...708..750S}. Similarly, the value of 0.08 is supposed to represent an rms variation from galaxy to galaxy, rather than an uncertainty in the mean.  In this expression, $\rho$ represents both the dark matter and stellar contributions to the mass density.  

\begin{figure}
\centering
\includegraphics[width=0.47\textwidth]{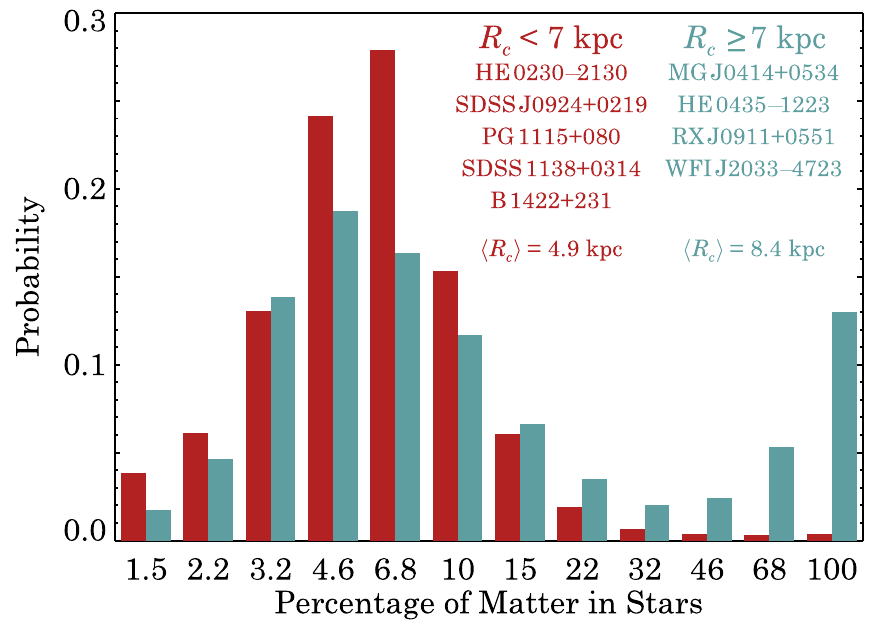}
\caption{As in Figure~\ref{fig:jointprob} for systems with known $z_l$, separated into two groups based on $R_c$.  \qlens\ and \rxelevenlens\ have been excluded.}
\label{fig:splitprob}
\end{figure}

The observations determine the most probable stellar fraction $S$, which is the fraction in stars of the total column density $C_\mathrm{tot}$ along the line of sight at impact parameter $R$.  We can integrate expressions (\ref{eq:stars}) and (\ref{eq:rho}) to obtain:
\begin{equation}
\frac{C_\mathrm{stars}}{C_\mathrm{tot}} = \frac{I_{\rm S0}}{\rho_0} \frac{\Gamma((\delta-1)/2)}{\Gamma((\alpha-1)/2)}\frac{\Gamma(\alpha/2)}{\Gamma(\delta/2)} \left(\frac{r_0}{R}\right)^{\delta - \alpha}
\end{equation}
where the only radial dependence is in the final factor, $R^{\alpha-\delta}$.  For the nominal values for $\delta$ and $\alpha$ listed above, this reduces to
\begin{equation}
S(R) = \frac{C_\mathrm{stars}}{C_\mathrm{tot}} \simeq 0.77\,  \frac{I_{\rm S0}}{\rho_0} \left(\frac{r_0}{R}\right)^{0.44}~~~~.
\label{eq:fracvsr}
\end{equation}
Therefore, for a range of mean impact parameters from $R_c = 3.9$ to 9.5 kpc, we expect the stellar fraction $S$ to vary by only a factor of $\sim$1.5 due to the dependence on the impact parameter.   Given that this is roughly the resolution of our logarithmic grid of a dozen values of $S$ and that our sample is modest in size, we would not expect our results to be sensitive to the range in $R_c$.

Nevertheless, we divided the 10 lens with known $z_l$ into two groups: those with $R_c < 7~\mathrm{kpc}$ and those with $R_c \geq 7~\mathrm{kpc}$.  The second group initially contained \rxelevenlens, but we removed it from the following analysis given the issues discussed in Section \ref{sec:rxj1131}.  We ran a joint analysis separately on these two groups, and the results are shown in Figure~\ref{fig:splitprob}.  As expected, the group with smaller $\langle R_c \rangle$ has a most probable stellar fraction $S$ that is larger than the group with larger $\langle R_c \rangle$.

To see how well this result agrees quantitatively with Equation~(\ref{eq:fracvsr}), we use a fitting function to determine the precise location of the peak of the probability distribution of each group.  The first group, with $\langle R_c \rangle = 4.9~\mathrm{kpc}$, peaks at $S_1 = 5.9\%\pm0.5\%$.  The second group,  with $\langle R_c \rangle = 8.4~\mathrm{kpc}$, peaks at $S_2 = 4.5\%\pm0.2\%$.  Based on the impact parameter ratio of 1.7, Equation~(\ref{eq:fracvsr}) predicts a stellar fraction ratio of 1.3. Somewhat remarkably, given the modest size of our samples, $S_1/S_2 = 1.3\pm0.13$. 

We also note the most likely stellar fraction of 100\% for \qlens\ (see Figure~\ref{fig:indivprobs}).  The lensing galaxy in this system is much closer than the others at $z_l = 0.04$ and consequently has a much smaller $R_c$ of 0.7 kpc.  It is the only system with a most likely stellar fraction of 100\%, and this is in qualitative agreement with expectations.  Quantitatively, it is larger than suggested by Equation~(\ref{eq:fracvsr}), but this may be due to either of Equations~(\ref{eq:stars}) or (\ref{eq:rho}) not being valid at such a small impact parameter.

\section{Summary}
\label{sec:summary}
We have analyzed 61 publicly available \chandra\ observations of 14 quadruply lensed quasars.  We extensively tested several methods to reduce and fit the \chandra\ data to obtain the best measurements of the individual X-ray fluxes of the quasar images.  As we have shown in our previous work \citep{2007ApJ...661...19P,2009ApJ...697.1892P}, the X-ray fluxes are a relatively clean measure of microlensing effects, unencumbered by source size considerations.  

The results of our data reduction and analysis were used in a Bayesian analysis of custom microlensing magnification maps which marginalized over all observational uncertainties as well as multiple observations of a lensed quasar.  Our analysis yields a most likely local stellar fraction of 6.8\% (i.e., a most likely dark matter fraction of 93.2\%) for the ensemble of lensing galaxies, integrated along the line of sight at a mean impact parameter of 6.6 kpc.  This is similar to the value of 5\% found by \citet{2009ApJ...706.1451M}, who studied flux ratios in the optical and assumed a source size of \ee{2.6}{15}~cm.  It is also consistent with the recent work of \citet{2011ApJ...731...71B}, which considered optical and infrared data and found dark matter fractions of 50\err{30}{40}\%, 80\err{10}{10}\%, and $\leq$50\% in \mglens, \sdssninelens, and \qlens, respectively.  Those authors performed a marginalization over the source size parameters in their analysis.  A distinct advantage of the work presented here is that our X-ray analysis is unencumbered by source-size considerations.

We formed two subsets of the lensing galaxies based on the mean impact parameters where their images formed and found that their most likely stellar fractions varied both qualitatively as expected---higher stellar fractions closer to the centers of the lensing galaxies---and quantitatively as expected.  In addition, we find a most likely stellar fraction of 100\% for \qlens, which has a mean impact parameter about an order of magnitude smaller than all of the other lens systems we studied.  

Our measurement of integrated stellar fraction as a function of impact parameter opens up the possibility of mapping out the dark matter content of lensing galaxies in a direct and straightforward manner, with minimal assumptions, based solely on high quality X-ray observations of lensed quasars. 

\acknowledgements 
The authors acknowledge and thank Alan Levine for several stimulating discussions. This research has made use of data obtained from the \chandra\ Data Archive  and software provided by the \chandra\ X-ray Center in the application packages CIAO and Sherpa. D.P.\ warmly thanks Craig Wheeler for his hospitality at UT Austin, where much of this work was performed.  J.A.B.\ and P.L.S.\ acknowledge support from NSF grant AST-0607601 and Chandra grant GO7-8099.


\begin{thebibliography}{}
\bibitem[Bate et al.(2011)]{2011ApJ...731...71B} Bate, N.~F., Floyd, D.~J.~E., Webster, R.~L., \& Wyithe, J.~S.~B.\ 2011, \apj, 731, 71 
\bibitem[Blackburne et al.(2006)]{2006ApJ...640..569B} Blackburne, J.~A., Pooley, D., \& Rappaport, S.\ 2006, \apj, 640, 569
\bibitem[Blackburne et al.(2011)]{2011ApJ...729...34B} Blackburne, J.~A., Pooley, D., Rappaport, S., \& Schechter, P.~L.\ 2011, \apj, 729, 34 
\bibitem[Blackburne et al.(2008)]{2008AJ....135..374B} Blackburne, J.~A., Wisotzki, L., \& Schechter, P.~L.\ 2008, \aj, 135, 374 
\bibitem[Cash(1979)]{1979ApJ...228..939C} Cash, W.\ 1979, \apj, 228, 939 
\bibitem[Chartas et al.(2009)]{2009ApJ...693..174C} Chartas, G., Kochanek, C.~S., Dai, X., Poindexter, S., \& Garmire, G.\ 2009, \apj, 693, 174
\bibitem[Chiba(2002)]{2002ApJ...565...17C} Chiba, M.\ 2002, \apj, 565, 17
\bibitem[Dalal \& Kochanek(2002)]{2002ApJ...572...25D} Dalal, N., \& Kochanek, C.~S.\ 2002, \apj, 572, 25 
\bibitem[Dickey \& Lockman(1990)]{1990ARA&A..28..215D} Dickey, J.~M., \& Lockman, F.~J.\ 1990, \araa, 28, 215 
\bibitem[Faure et al.(2004)]{2004A&A...428..741F} Faure, C., Alloin, D., Kneib, J.~P., \& Courbin, F.\ 2004, \aap, 428, 741 
\bibitem[Freeman et al.(2001)]{2001SPIE.4477...76F} Freeman, P., Doe, S., \& Siemiginowska, A.\ 2001, \procspie, 4477, 76 
\bibitem[Gavazzi et al.(2007)]{2007ApJ...667..176G} Gavazzi, R., Treu, T., Rhodes, et al.\ 2007, \apj, 667, 176 
\bibitem[Giacconi et al.(2001)]{2001ApJ...551..624G} Giacconi, R., et al.\ 2001, \apj, 551, 624 
\bibitem[Kneib et al.(1998)]{1998A&A...339L..65K} Kneib, J.-P., Alloin, D., \& Pello, R.\ 1998, \aap, 339, L65 
\bibitem[Kochanek et al.(2007)]{2007ASPC..371...43K} Kochanek, C.~S., Dai, X., Morgan, C., Morgan, N., \& Poindexter, S.~C., G.\ 2007, Statistical Challenges in Modern Astronomy IV, 371, 43 
\bibitem[Kochanek \& Dalal(2004)]{2004ApJ...610...69K} Kochanek, C.~S., \& Dalal, N.\ 2004, \apj, 610, 69 
\bibitem[Kochanek et al.(2006)]{2006ApJ...640...47K} Kochanek, C.~S., Morgan, N.~D., Falco, E.~E., et al.\ 2006, \apj, 640, 47 
\bibitem[Koopmans(2006)]{2006EAS....20..161K} Koopmans, L.~V.~E.\ 2006, EAS Publications Series, 20, 161
\bibitem[Koopmans et al.(2009)]{2009ApJ...703L..51K} Koopmans, L.~V.~E., Bolton, A., Treu, T., et al.\ 2009, \apjl, 703, L51
\bibitem[Kroupa(2001)]{2001MNRAS.322..231K} Kroupa, P.\ 2001, \mnras, 322, 231
\bibitem[Li et al.(2004)]{2004ApJ...610.1204L} Li, J., Kastner, J.~H., Prigozhin, G.~Y., et al.\ 2004, \apj, 610, 1204 
\bibitem[Magain et al.(1988)]{1988Natur.334..325M} Magain, P., Surdej, J., Swings, J.-P., Borgeest, U., \& Kayser, R.\ 1988, \nat, 334, 325 
\bibitem[Mao \& Schneider(1998)]{1998MNRAS.295..587M} Mao, S., \& Schneider, P. \ 1998, MNRAS, 295, 587
\bibitem[Mediavilla et al.(2009)]{2009ApJ...706.1451M} Mediavilla, E., Mu\~{n}oz, J.~A., Falco, E., et al.\ 2009, \apj, 706, 1451 
\bibitem[Metcalf \& Madau(2001)]{2001ApJ...563....9M} Metcalf, R.~B., \& Madau, P.\ 2001, \apj, 563, 9 
\bibitem[Metcalf \& Zhao(2002)]{2002ApJ...567L...5M} Metcalf, R.~B., \& Zhao, H.\ 2002, \apjl, 567, L5 
\bibitem[Morgan et al.(2008)]{2008ApJ...689..755M} Morgan, C.~W., Kochanek, C.~S., Dai, X., Morgan, N.~D., \& Falco, E.~E.\ 2008, \apj, 689, 755 
\bibitem[Morgan et al.(2004)]{2004AJ....127.2617M} Morgan, N.~D., Caldwell, J.~A.~R., Schechter, P.~L., et al.\ 2004, \aj, 127, 2617 
\bibitem[Nelder \& Mead(1965)]{nm65}Nelder, J.~A. \& Mead, R.\ 1965, Comput.\ J, 7, 308
\bibitem[Pooley et al.(2007)]{2007ApJ...661...19P} Pooley, D., Blackburne, J.~A., Rappaport, S., \& Schechter, P.~L.\ 2007, \apj, 661, 19 
\bibitem[Pooley et al.(2006)]{2006ApJ...648...67P} Pooley, D., Blackburne, J.~A., Rappaport, S., Schechter, P.~L., \& Fong, W.-f.\ 2006, \apj, 648, 67 
\bibitem[Pooley et al.(2009)]{2009ApJ...697.1892P} Pooley, D., Rappaport, S., Blackburne, J., et al.\ 2009, \apj, 697, 1892 
\bibitem[Schechter \& Wambsganss(2002)]{2002ApJ...580..685S} Schechter, P.~L., \& Wambsganss, J.\ 2002, \apj, 580, 685 
\bibitem[Schechter \& Wambsganss(2004)]{2004IAUS..220..103S} Schechter, P.~L., \& Wambsganss, J.\ 2004, in IAU Symp.\ 220, Dark Matter in Galaxies, ed.\ S.~D.\ Ryder et al.\ (San Francisoc, CA: ASP), 103 
\bibitem[Schwab et al.(2010)]{2010ApJ...708..750S} Schwab, J., Bolton, A.~S., \& Rappaport, S.~A.\ 2010, \apj, 708, 750
\bibitem[Treu et al.(2006)]{2006ApJ...640..662T} Treu, T., Koopmans, L.~V., Bolton, A.~S., Burles, S., \& Moustakas, L.~A.\ 2006, \apj, 640, 662
\bibitem[Wambsganss(1990)]{1990PhDT.......180W} Wambsganss, J.\ 1990, PhD~thesis Ludwig-Maximilians-Universit\:{a}t Munich
(preprint MPA 550)
\bibitem[Wambsganss(1999)]{1999JCoAM.109..353W} Wambsganss, J.\ 1999, J.\ Comput.\ Appl.\ Math., 109, 353
\bibitem[Wambsganss \& Paczy\'nski(1992)]{1992ApJ...397L...1W} Wambsganss, J., \& Paczy\,nski, B.\ 1992, \apjl, 397, L1 
\bibitem[Wambsganss et al.(1990)]{1990ApJ...352..407W} Wambsganss, J., Paczynski, B., \& Katz, N.\ 1990, \apj, 352, 407 
\bibitem[Witt et al.(1995)]{1995ApJ...443...18W} Witt, H., Mao, S., \& Schechter, P.~L.\ 1995, \apj, 443, 18 
\bibitem[Wyithe et al.(2000)]{2000MNRAS.312..843W} Wyithe, J.~S.~B., Webster, R.~L., \& Turner, E.~L.\ 2000, \mnras, 312, 843 

\end{thebibliography}
\end{document}